\def\mycolor{black}
\def\webDOI{http://dx.doi.org/}

\documentclass[review,1p,times]{amsart}

\usepackage{epsfig}

\usepackage{amssymb}
\usepackage{amsthm}
\usepackage{amsmath,stmaryrd}
\usepackage{color}
\usepackage[pdftex, pdfborderstyle={/S/U/W 1}]{hyperref}
\usepackage{url}
\usepackage{cite}
\usepackage{nccmath}
\usepackage{cancel}

\usepackage{lineno}

\newcommand{\fref}[1]{Fig.~\ref{#1}}  
\newcommand{\eref}[1]{Eq.~\eqref{#1}} 
\newcommand{\sref}[1]{Sect.~\ref{#1}} 
\newcommand{\PBS}[1]{\let\temp=\\#1\let\\=\temp}

\newcommand{\aref}[1]{Appendix \ref{#1}}

\newcommand{\ci}{\mathrm{i}}
\newcommand{\supp}{\operatorname{supp}}

\newcommand{\varCset}{\mathcal{C}}
\newcommand{\varSset}{\mathcal{S}}

\newcommand{\itr}{{\sf T}}
\newcommand{\xgj}{x}
\newcommand{\ygj}{y}

\newcommand{\vgj}{v}

\newcommand{\yg}{{\boldsymbol\ygj}}

\newcommand{\Vg}{{\boldsymbol V}}

\newcommand{\vg}{{\boldsymbol\vgj}}

\newcommand{\specij}{W}

\newcommand{\eigl}{\omega}

\newcommand{\II}{{\boldsymbol I}}

\newcommand{\Go}{\mathrm{O}}
\newcommand{\po}{\mathrm{o}}
\newcommand{\varXset}{{\mathcal X}}

\newcommand{\xg}{{\boldsymbol\xgj}}
\newcommand{\vpj}{p}

\newcommand{\vp}{{\boldsymbol\vpj}}

\newcommand{\bzero}{{\bf 0}}					
\newcommand{\kgj}{k}
\newcommand{\kg}{{\boldsymbol\kgj}}
\newcommand{\hkg}{{\hat\kg}}
\newcommand{\pg}{{\boldsymbol p}}
\newcommand{\hpg}{{\hat\pg}}

\newcommand{\wg}{\omega}					
\newcommand{\pres}{p}
\newcommand{\roi}{\varrho}					
\newcommand{\cel}{c}
\newcommand{\Cel}{C}

\newcommand{\domain}{{\mathcal O}}			

\newcommand{\id}{d}
\newcommand{\iD}{\mathrm{D}}
\newcommand{\dd}{{\mathrm d}}				
\newcommand{\dconv}[1]{\frac{\dd{#1}}{\dd t}}
\newcommand{\bnabla}{{\boldsymbol\nabla}}		
\newcommand{\Dx}{{\bf D}}				
\newcommand{\Dxx}{\Dx_\xg}
\newcommand{\Dxt}{\Dx_\xt}

\newcommand{\Dt}{\iD_t}	
\newcommand{\Wigner}{W}				
\newcommand{\TF}[1]{\widehat{#1}}				
\newcommand{\iexp}{\operatorname{e}}			

\newcommand{\cjg}[1]{\overline{#1}}
\newcommand{\adj}[1]{{#1}^*}
\newcommand{\norm}[1]{|#1|}
\newcommand{\normu}[1]{\left|#1\right|}
\newcommand{\esp}[1]{{\mathbb E}\left\{#1\right\}}
\newcommand{\scal}[1]{\left\langle#1\right\rangle}
\newcommand{\scald}[1]{\left(#1\right)_{L^2}}

\newcommand{\Rset}{\mathbb{R}}			
\newcommand{\Cset}{\mathbb{C}}

\newcommand{\vC}{\Omega}
\newcommand{\vL}{L}
\newcommand{\vH}{H}

\newcommand{\demi}{\frac{1}{2}}	      

\newcommand{\fu}{\phi}
\newcommand{\fv}{\psi}
  
\newcommand{\dir}{\delta}                       
\newcommand{\obsj}{P}

\newcommand{\coroij}{R}
\newcommand{\coro}{{\boldsymbol\coroij}}        

\newcommand{\indL}{2}

\newcommand{\iref}{0}
\newcommand{\escale}{\varepsilon}

\newcommand{\Wignere}{\Wigner_\escale}
\newcommand{\fue}{\fu_\escale}
\newcommand{\fve}{\fv_\escale}
\newcommand{\vref}{\vg_\iref}
\newcommand{\Vref}{\Vg_\iref}
\newcommand{\vper}{\vg_1}
\newcommand{\Vper}{\Vg_1}
\newcommand{\rref}{\roi_\iref}
\newcommand{\pref}{\pres_\iref}
\newcommand{\cref}{\cel_\iref}
\newcommand{\Cref}{\Cel_\iref}
\newcommand{\cper}{\chi}
\newcommand{\vpot}{\phi}
\newcommand{\vpote}{\vpot_\escale}
\newcommand{\xt}{{\boldsymbol s}}
\newcommand{\xte}{\frac{\xt}{\escale}}
\newcommand{\yt}{{\boldsymbol\tau}}
\newcommand{\kw}{{\boldsymbol\xi}}
\newcommand{\pw}{{\boldsymbol\eta}}
\newcommand{\apar}{a}

\newcommand{\acoeff}{\TF{\alpha}_1}
\newcommand{\aacoeff}{\TF{\alpha}_2}
\newcommand{\vray}{\vg_g}

\newcommand{\action}{\mathcal{A}}

\newcommand{\sequence}[1]{(#1)}

\newcommand{\sig}[1]{\hat{#1}}

\newcommand{\dscatij}{\sigma}

\newcommand{\tscati}{\Sigma}

\newcommand{\emphes}[1]{\textcolor{black}{#1}}

\newcommand{\alert}[1]{\textcolor{black}{#1}}

\begin{document}




\title[Multiple scattering of waves in a random flow]{Kinetic modeling of multiple scattering of acoustic waves in randomly heterogeneous flows}


\author[J.-L. Akian]{Jean-Luc Akian}
\address[J.-L. Akian]{ONERA--The French Aerospace Lab, France}
\email{jean-luc.akian@onera.fr}

\author[\'E. Savin]{\'Eric Savin}
\address[\'E. Savin]{ONERA--The French Aerospace Lab, France}
\thanks{Corresponding author: \'E. Savin, ONERA--The French Aerospace Lab, 6 chemin de la Vauve aux Granges, FR-91123 Palaiseau cedex, France (Eric.Savin@onera.fr).}
\email{eric.savin@onera.fr}



\begin{abstract}
We study the propagation of sound waves in a three-dimensional, infinite ambient flow with weak random fluctuations of the mean particle velocity and speed of sound.  We more particularly address the regime where the acoustic wavelengths are comparable to the correlation lengths of the weak inhomogeneities--the so-called weak coupling limit. The analysis is carried on starting from the linearized Euler equations and the convected wave equation with variable density and speed of sound, which can be derived from the nonlinear Euler equations. We use a multi-scale expansion of the Wigner distribution of a velocity potential associated to the waves to derive a radiative transfer equation describing the evolution of the angularly resolved wave action in space/time phase space. The latter experiences convection, refraction and scattering when it propagates through the heterogeneous ambient flow, although the overall wave action is conserved. The convection and refraction phenomena are accounted for by the convective part of the transport equation and depend on the smooth variations of the ambient quantities. The scattering phenomenon is accounted for by the collisional part of the transport equation and depends on the cross-power spectral densities of the fluctuations of the ambient quantities at the wavelength scales. The refraction, phase shift, spectral broadening, and multiple scattering effects of the high-frequency regimes described in various previous publications are thus encompassed by the proposed model. The overall derivation is based on the interpretation of spatial-temporal Wigner transforms in terms of semiclassical operators \emphes{in their standard quantization}.
\end{abstract}

\keywords{Linearized Euler equations, Acoustic waves, Kinetic model, Transport equation, Radiative transfer}

\date{\today}


\maketitle

\section{\textcolor{\mycolor}{Introduction and summary}}

\subsection{Modeling of acoustic wave propagation in random flow}\label{sec:}

The study of multiply-scattered acoustic waves in heterogeneous, unsteady flows has relevance to atmospheric and ocean acoustics, infrasound propagation, acoustics of turbulent flows, or even astrophysics, among other examples; see \cite{CAR83,JEN11,OST16,PIE89} and references therein. Applications concern acoustic remote sensing and tomography in the atmosphere and ocean, noise emission by nozzles and exhaust pipes, localization of acoustic sources, or prediction and reduction of sound waves from infrastructures for instance. The analysis of sound wave propagation in media with random perturbations of the speed of sound has been well developed in the past; see \emph{e.g.} \cite{ISH78,KAR64,TAT61}. In jet shear layers for example the speed of sound is likely to exhibit large fluctuations on very short propagation paths, producing a spectral broadening of the high-frequency tones propagating through them \cite{CAR83}. However, in the atmosphere or in the ocean the sound pressure is also \alert{influenced} by inhomogeneities of the density and current velocity as argued in \emph{e.g.} \cite{OST16}. As an acoustic wave propagates through a heterogeneous, unsteady ambient flow it undergoes convection, refraction, scattering, and absorption. Convection is responsible for the Doppler effect which shifts the typical frequency of the waves. Refraction corresponds to a change of directivity of a sound beam induced by the mean velocity gradient of the flow. Scattering by turbulence causes a redistribution of the acoustic energy among different wave numbers and frequencies resulting in spectral and directional broadening of the waves at sufficiently high frequencies. Absorption of the waves is seen as their energy is partly transferred to the heterogeneous ambient flow in the propagation process; see \cite{CAN76,GUE85}.

Various models of acoustic wave propagation in unsteady, heterogeneous flows have been developed to analyze these phenomena. Iterative perturbation expansions for weakly random inhomogeneities of the ambient quantities were considered in~\cite{FRI68,WEN71} following the seminal developments of Karal \& Keller~\cite{KAR64}. Single-scattering approximations, or Born approximations were considered in several earlier works alike \cite{BLO46a,GUE85,KRA53,LIG53}. The so-called sound scattering cross-section of a plane wave experiencing scattering from wind velocity and temperature fluctuations in a small region of inhomogeneous fluid has also been proposed as a mean to characterize single-scattering in a turbulent medium \cite{KRA53,MON62}. Geometric acoustics or ray tracing is a popular approach to model sound propagation in moving media, though \cite{BLO46a,CAN77,GRO55,HAY68,KOR53,MIL21,PIE90,THO65}. It is a high-frequency approximation whereby the sound field is expanded in a power series of the small acoustic wavelength. Although it is very efficient for simulating long range propagation of intense noise (as sonic boom or explosions for example), it is limited in several configurations such as shadow zones or caustics. It is in addition not able to handle scattering by fine structures or turbulence and generally fails when multiple scattering occurs. 
Parabolic approximations are classically considered to work beyond ray acoustics and their shortcomings but have usually been limited to low Mach numbers; see for example \cite{OST94} or \cite{OST16} for a review of their actual developments. This limitation reduces their relevance to applications in atmospheric propagation and precludes applications in turboengine noise propagation. However a combination of ray acoustics and parabolic approximation for larger Mach numbers may be envisaged, as proposed in \cite{COU08}.

Here we rather focus on transport and radiative transfer models which describe the mesoscopic regime of wave propagation. The wavelength is comparable to the characteristic length of the inhomogeneities, typically a correlation length in the actual flow (hereafter referred to as the fast scale). This regime corresponds to a situation of strong interaction between waves and random heterogeneities which cannot be addressed by usual homogenization or iterative techniques. Also it considers large propagation distances compared to the wavelength, and weak amplitudes of the random perturbations in the actual flow with respect to a--possibly heterogeneous--ambient flow varying at a length scale \textcolor{\mycolor}{(the slow scale)} one order of magnitude larger than the wave/correlation lengths. This corresponds to the so-called weak coupling limit as defined in the dedicated literature (see for example \cite{FOU07}), whereby an explicit separation of scales can be invoked. The analysis developed in \cite{AKI12,BAL02,BAL05,BAY14,BOR19,BRA02,FAN01,GUO99,POW05,RYZ96} is based on the use of a Wigner transform of the wave field, the high-frequency, non-negative limit of which characterizing its angularly resolved energy density. It can be made mathematically rigorous as in~\cite{AKI12,GER97,LIO93} ignoring however the influence of random inhomogeneities, except for some particular situations~\cite{ERD00,LUK07}.

In this research, two kinds of perturbations of the ambient quantities are considered in order to assess their influence on the transport regime of acoustic waves in unsteady, heterogeneous flows. The first one is related to the speed of sound, of which variations are typically induced by variations of the ambient temperature and humidity in the atmosphere. The second one is related to the particle velocity of the flow, of which variations are typically induced by the turbulent structures of the former. More specifically, a formal radiative transfer model is developed to explain the propagation of multiply-scattered acoustic waves in such a flow. This model describes the evolution of the \emph{wave action} \cite{WHI74} in phase space in terms of the Wigner measure of the wave fields. The analysis starts from a second-order wave equation and introduces the spatio-temporal Wigner transform of the acoustic waves. It follows the techniques used by Bal~\cite{BAL05}, Akian~\cite{AKI12} or Baydoun \emph{et al.} \cite{BAY14}. However, as opposed to~\cite{BAL05} it considers convected acoustic waves \emphes{and the standard quantization of the Wigner transform and associated pseudo-differential calculus}, and as opposed to~\cite{AKI12} it considers the influence of random perturbations in the weak coupling regime. Our derivation generalizes the earlier developments of Howe \cite{HOW73} in that it obtains in a systematic way the high-frequency kinetic limit of the wave equation with explicit, general perturbations of the ambient quantities. This limit of course encompasses the results of \cite{HOW73}. The developments of \cite{FAN01} rely on the same mathematical tools, however they consider time-independent ambient quantities and the first-order linearized Euler equations rather than the second-order convected wave equation addressed below. The developments of \cite{BAL02} consider a time-dependent flow velocity as done here. \emphes{At last the analysis in \cite{BOR19} considers a constant mean velocity of the ambient flow and acoustic waves in a forward-scattering regime of propagation}.

\subsection{Summary of the main results}

We now summarize our main results. We consider acoustic waves in an inhomogeneous, unsteady ambient flow and their multiple scattering induced by random perturbations of the flow characteristics. We also want to account for the space- and time-dependence of these perturbations, and address the regime where the leading wavelength is comparable to their (small) correlation length. This setting ensures maximum interactions with the acoustic waves, a necessary condition if we want to probe with waves in random media as for applications of correlation-based imaging techniques \cite{AMM13,BOR05,GAR09} we ultimately have in mind. It defines the high-frequency range terminology we shall use throughout the paper.

As a sound wave propagates in a \emph{quiescent}, randomly perturbed medium with an incident wave vector $\smash{\kg'}$, it can be scattered at any time $t$ and position $\xg$ into any direction $\smash{\hkg}$ and wave vector $\kg$ (such that $\smash{\hkg}=\kg/|\kg|$). Therefore it is relevant to consider an angularly resolved energy density $\Wigner(\xg,t;\kg)$ for this wave, defined in phase space. In~\cite{RYZ96,BAL05} it is shown that energy conservation takes the form of a radiative transfer equation:
\begin{equation}\label{eq:RTE-acou}
\partial_t\Wigner(\xg,t;\kg)+\{\lambda_\pm,\Wigner\}+\tscati(\xg,\kg)\Wigner(\xg,t;\kg)=\int\dscatij(\xg;\kg|\kg')\Wigner(\xg,t;\kg')\dd\kg'\,,
\end{equation}
where $\smash{\lambda_\pm}(\xg,\kg)=\pm\smash{\Cref}(\xg)\norm{\kg}$ is the frequency of the waves at $\xg$ with wave vector $\pm\kg$, $\Cref(\xg)$ is the sound speed in the unperturbed background medium, and $\{a,b\}=\bnabla_\kg a\cdot\bnabla_\xg b-\bnabla_\xg a\cdot\bnabla_\kg b$ stands for the usual Poisson bracket. The kernel $\dscatij(\xg;\kg|\kg')$ is the rate of conversion of energy with wave vector $\smash{\kg'}$ into energy with wave vector $\kg$ at position $\xg$--the so-called scattering cross-section. The total scattering cross-section $\tscati$ is:
\begin{displaymath}
\tscati(\xg,\kg)=\int\dscatij(\xg;\kg|\kg')\dd\kg'\,,
\end{displaymath}
such that the radiative transfer equation is conservative because the former relationship yields:
\begin{displaymath}
\iint\Wigner(\xg,t;\kg)\dd\kg\dd\xg=\operatorname{Const}
\end{displaymath}
for all times. The scattering cross-section is explicitly determined by the power spectral density of the perturbations~\cite{RYZ96,BAL05}. \alert{The wave frequency is also kept constant in the scattering processes described by the radiative transfer equation (\ref{eq:RTE-acou})}. This model remains valid when the waves are scattered by randomly distributed discrete inclusions, in which case the scattering cross-section is the cross-section of a single inclusion multiplied by their density. Here we only consider continuous random inhomogeneities.

For a \emph{moving} ambient medium, we must in addition take account of the convective effects of the waves, and time-dependence of the ambient quantities. As a sound wave propagates in the flow with wave vector $\smash{\kg'}$ and frequency $\smash{\wg'}$, it can now be scattered at any time $t$ and position $\xg$ into any direction $\smash{\hkg}$, wave vector $\kg$, and frequency $\wg$. It is now relevant to consider a directionally and frequency resolved energy density $\Wigner(\xg,t;\kg,\wg)$ for this wave, defined in the phase space of the position-time space. Then it is shown in \sref{sec:RTE} that the radiative transfer equation (\ref{eq:RTE-acou}) takes the following form for a moving medium:
\begin{equation}\label{eq:RTE-acout}
\partial_t\action(\xg,t;\kg) + \{\lambda_\pm,\action\}+\tscati(\xg,t;\kg)\action(\xg,t;\kg)=\int\dscatij(\xg,t;\kg|\kg')\action(\xg,t;\kg')\dd\kg'\,,
\end{equation}
where the Doppler-shifted frequency of the waves is actually given by $\wg=\smash{\lambda_\pm(\xg,t;\kg)}\\=\smash{\Vref(\xg,t)\cdot\kg\pm\Cref(\xg,t)\norm{\kg}}$, and $\Cref(\xg,t)$ and $\Vref(\xg,t)$ are the speed of sound and particle velocity of the unperturbed ambient flow, respectively.  Also $\action(\xg,t;\kg)=\smash{\frac{\rref(\xg,t)}{\Cref(\xg,t)}}\norm{\kg}\Wigner(\xg,t;\kg,\lambda_\pm)$ is the wave action \cite[Chapter 11]{WHI74}, $\rref(\xg,t)$ being the density of the unperturbed ambient flow. The differential scattering cross-section $\smash{\dscatij(\xg,t;\kg|\kg')}$ is again explicitly determined in terms of the power spectral density of the random inhomogeneities, which accounts for perturbations of the speed of sound, the particle velocity, and their possible correlations. This model implies that the ambient flow quantities and their perturbations have the following contributions to the wave dynamics:
\begin{itemize}
\item The variations of the ambient quantities $\Cref$ and $\Vref$ at the slow scale contribute only to the left-hand side of the radiative transfer equation (\ref{eq:RTE-acout}). They basically characterize the \emph{group velocity} $\smash{\vray^\pm}:=\smash{\bnabla_\kg\lambda_\pm}$ of the waves in the transport regime, and account for both the phase shift (change of direction) and spectral broadening (change of frequency) effects. \alert{\sref{sec:Transport} explains how this dynamics, which is the classical geometric acoustics or ray tracing approach of sound propagation in a moving medium already outlined above, is recovered from our theoretical analyses}.
\item The perturbations of the ambient quantities primarily contribute to the right-hand side of the radiative transfer equation in terms of their cross-power spectral densities. This collisional kernel describes how high-frequency waves are continuously scattered by the flow inhomogeneities at the fast scale, which is also their (small) wavelength. It thus models the multiple-scattering effects \alert{analyzed in detail in \sref{sec:RTE}. Here we generalize the radiative transfer model \eqref{eq:RTE-acou} to a moving medium with possibly time-varying characteristics of the background flow using revisited rules of pseudo-differential calculus. This section contains the main novelties introduced in this paper}.  
\item The total scattering cross-section $\smash{\tscati(\xg,t;\kg)}$ describes both the overall acoustic energy that is scattered into all other directions and frequencies by the multiple scattering process, and the acoustic energy that is irreversibly transferred to the ambient flow. \alert{In \sref{second-corre-sec} more particularly we also generalize the expressions of the \emph{time-independent} differential and total scattering cross-sections $\smash{\dscatij(\xg;\kg|\kg')}$ and $\tscati(\xg;\kg)$, respectively, in \eref{eq:RTE-acou} for a \emph{quiescent} medium, to their \emph{time-dependent} counterparts $\smash{\dscatij(\xg,t;\kg|\kg')}$ and $\tscati(\xg,t;\kg)$, respectively, in \eref{eq:RTE-acout} for a \emph{moving} medium. This is also a new result of the paper}.
\end{itemize}

\begin{figure}
\alert{\includegraphics[scale=.33]{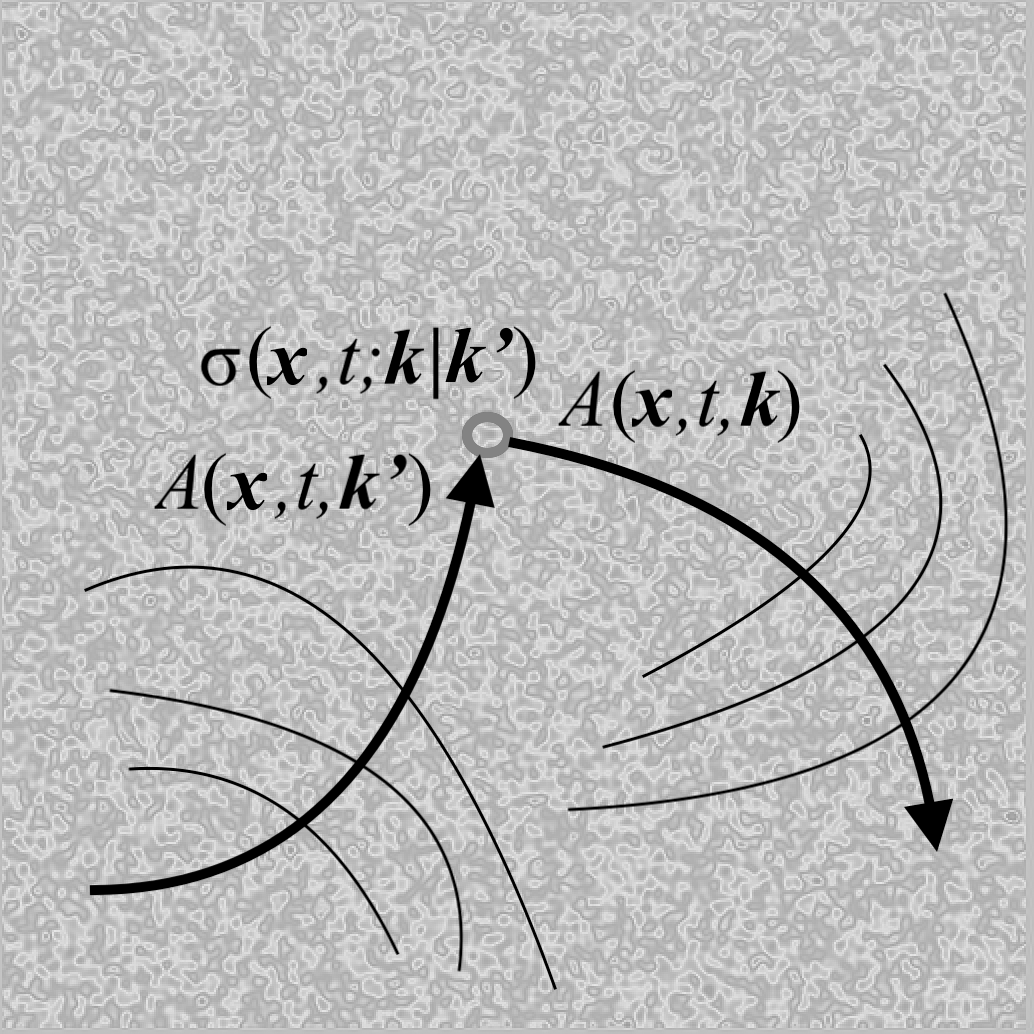}
\caption{Differential scattering cross-section $\dscatij(\xg,t;\kg|\kg')$ for scattering from wave action $\action(\xg,t;\kg')$ with wave vector and frequency $\kg'$ to wave action $\action(\xg,t;\kg)$ with wave vector and frequency $\kg$, at the time $t$ and position $\xg$ shown by the grey circle. The random fluctuations of the ambient medium sketched in light grey have small amplitudes and a small correlation length comparable to the acoustic wavelength.}\label{fg:scat-x-section}}
\end{figure}

\alert{This mechanism of conversion of the wave actions by scattering on inhomogeneities as described by their differential scattering cross-section $\smash{\dscatij(\xg,t;\kg|\kg')}$ is illustrated on \fref{fg:scat-x-section}. The forward wave action $\action(\xg,t;\kg')$ with wave vector and frequency $\kg'$ is converted to the wave action $\action(\xg,t;\kg)$ with wave vector and frequency $\kg$ with a rate $\dscatij(\xg,t;\kg|\kg')$ at time $t$ and position $\xg$ depending only on the second-order statistics (cross-power spectral densities) of the random fluctuations of the background medium. The paths of the wave actions are curved because of the variations of the ambient quantities $\Cref$ and $\Vref$ at the slow scale. At last the typical size (correlation length) of the random fluctuations sketched in light grey on \fref{fg:scat-x-section} is small but comparable to the wavelength}.

Of course \eref{eq:RTE-acout} can be recast as a conservative radiative transfer equation (\ref{eq:RTE-acou}) for a vanishing particle velocity of the ambient flow and time-independent ambient quantities. This is shown in \sref{sec:RTE} alike. We conclude this introduction by noting that kinetic models for first-order hyperbolic systems in media with time-dependent random perturbations have already been studied in \cite{BRA02} in a general mathematical framework.

\subsection{Outline}

The rest of the paper is organized as follows. In \sref{sec:HFwaves} we introduce the basic physical framework and notations used throughout. We more particularly focus on the characterization of sound propagation in a flow by a convected wave equation. \emphes{The spatio-temporal Wigner transform and the formal mathematical tools used in our analyses are reviewed in \sref{sec:PDO}. Here the relevance of considering a Wigner transform and its non-negative limit measure for the analysis of multiple scattering phenomena in the high-frequency range is emphasized. This limit is simply the energy density $\Wigner$ introduced above. We also stress that the quantization of the semi-classical operators and Wigner transform we use here is different from the one classically invoked in \emph{e.g.} \cite{BAL02,BAL05,BAY14,FAN01,RYZ96}, which we argue clarifies the derivation of their relevant properties}. The corresponding transport model is then derived in detail in \sref{sec:Transport} ignoring the influence of random inhomogeneities in a first step. The main contribution of the paper is \sref{sec:RTE} which outlines the extension of the previous transport model to account for random perturbations of the speed of sound and particle velocity in the ambient flow. A radiative transfer equation is obtained in the most general case. Its collisional kernel is explicitly described in terms of the correlation structure of the random perturbations. In this respect, it should be noted that the proposed theory requires a full characterization of the power spectral densities of these perturbations (assumed to be statistically homogeneous at the small wavelength scale), but no other statistical information. Some conclusions and perspectives are finally drawn in \sref{sec:CL}.

\section{Model of sound propagation in unsteady inhomogeneous flow}\label{sec:HFwaves}

In this section we establish the model of sound wave propagation in an unsteady inhomogeneous flow we are interested in for the derivation of the multiple-scattering kinetic (transport) model that will be detailed in the subsequent parts. The primary objective is to introduce the main notations that will be used throughout the paper.

\subsection{Linearized Euler equations about an unsteady inhomogeneous flow}\label{sec:HFwaves-basics}

The full non-linear Euler equations for an ideal fluid flow in the absence of friction, heat conduction, or heat production are:
\begin{equation}
\begin{split}
\dconv{\roi} &+\roi\bnabla_\xg\cdot\vg=0\,, \\
\dconv{\vg} &+\frac{1}{\roi}\bnabla_\xg\pres=\bzero\,, \\
\dconv{s} &=0\,,
\end{split}
\end{equation}
where $\roi$ is the fluid density, $\vg$ is the particle velocity, $s$ is the specific entropy, and $\pres$ is the thermodynamic pressure given by the equation of state $\pres=\pres(\roi,s)$. Also:
\begin{displaymath}
\dconv{}=\frac{\partial}{\partial t}+\vg\cdot\bnabla_\xg
\end{displaymath}
is the usual convective derivative following the particle paths. This shows that the flow is isentropic (\emph{i.e.} each fluid particle has constant entropy but different particles may have different entropy), and by the equation of state:
\begin{displaymath}
\dconv{\pres}=\cel^2\dconv{\roi}\,,\quad\cel^2(\roi,s)=\left(\frac{\partial\pres}{\partial\roi}\right)_s\,,
\end{displaymath}
where $\cel$ is the speed of sound.

Linearized acoustics equations arise from the previous conservation equations when its variables are expressed as sums of ambient quantities pertaining to the background flow (subscript $\iref$), and lower-order acoustic perturbations (primed quantities):
\begin{displaymath}
\begin{split}
\roi(\xg,t) &=\rref(\xg,t)+\roi'(\xg,t)\,, \\
\vg(\xg,t) & = \vref(\xg,t)+\vg'(\xg,t)\,, \\
s(\xg,t) & =s_\iref(\xg,t)+s'(\xg,t)\,,\\ 
\pres(\xg,t) & = \pref(\xg,t)+\pres'(\xg,t)\,.
\end{split}
\end{displaymath}
In such a manner, the primed quantities satisfy the linearized Euler equations:
\begin{equation}
\begin{split}
\dconv{\roi'} &+\roi'\bnabla_\xg\cdot\vref+\bnabla_\xg\cdot(\rref\vg')=0\,, \\
\dconv{\vg'} &+\vg'\cdot\bnabla_\xg\vref+\frac{1}{\rref}\bnabla_\xg\pres'-\frac{\roi'}{\rref^2}\bnabla_\xg\pref=\bzero\,, \\
\dconv{s'} &+\vg'\cdot\bnabla_\xg s_\iref=0\,,
\end{split}
\end{equation}
where here and in the remaining of the paper:
\begin{equation}\label{eq:dconv0}
\dconv{}=\frac{\partial}{\partial t}+\vref\cdot\bnabla_\xg\,,
\end{equation}
that is, the convective derivative within the ambient flow. In addition, one has from the linearized equation of state:
\begin{displaymath}
\pres'=\cref^2\roi'+\left(\frac{\partial\pres}{\partial s}\right)_{\rref} s'\,,
\end{displaymath}
where $\cref(\xg,t)>0$ is the sound speed in the ambient flow. In turn, the ambient variables satisfy:
\begin{equation}\label{eq:ambient-cons}
\begin{split}
\dconv{\rref} &+\rref\bnabla_\xg\cdot\vref=0\,, \\
\dconv{\vref} &+\frac{1}{\rref}\bnabla_\xg\pref=\bzero\,, \\
\dconv{s_\iref} &=0\,,
\end{split}
\end{equation}
together with the following relations from the equation of state applicable to the ambient flow:
\begin{equation}\label{eq:ambient-state}
\begin{split}
\bnabla_\xg\pref &=\cref^2\bnabla_\xg\rref+\left(\frac{\partial\pres}{\partial s}\right)_{\rref}\bnabla_\xg s_\iref\,, \\
\dconv{\pref} &=\cref^2\dconv{\rref}+\left(\frac{\partial\pres}{\partial s}\right)_{\rref}\dconv{s_\iref}\,.
\end{split}
\end{equation}

\subsection{Acoustic wave equation}

We now introduce a velocity quasi-potential $\vpot$ such that \cite{BLO46a,PIE90}:
\begin{equation}\label{eq:vpot}
\pres'=-\rref\dconv{\vpot}\,,\quad\vg'=\bnabla_\xg\vpot+\Go(L^{-1})+\Go(T^{-1})\,,
\end{equation}
where $L$ is a length scale over which the ambient quantities have significant spatial variations (the outer scale of turbulence), and $T$ is an associated time scale. These scales are assumed to be much larger than the corresponding ones for the acoustic disturbances, \emph{i.e.} the primed quantities. Then it is shown in \cite{BLO46a} that for a steady, irrotational ambient flow which is in addition homentropic (the entropy is constant and the same for all particles) one has:
\begin{equation}\label{eq:conv-wave}
\bnabla_\xg\cdot(\rref\bnabla_\xg\vpot)-\rref\dconv{}\left(\frac{1}{\cref^2}\dconv{\vpot}\right)=0\,,
\end{equation}
where $\smash{\dconv{}}$ is given by \eref{eq:dconv0}. It is a convected wave equation with the ambient inhomogeneous and steady particle velocity, density and sound speed $\vref(\xg)$, $\rref(\xg)$ and $\cref(\xg)$, respectively. In \cite{PIE90} it is shown that this model approximately extends to unsteady ambient flow up to second and higher-order terms with respect to the first-order terms pertaining to the ambient flow. Thus the above convected wave equation (\ref{eq:conv-wave}) with \emph{time-dependent} ambient quantities $\vref(\xg,t)$, $\rref(\xg,t)$, and $\cref(\xg,t)$ is the model of sound wave propagation considered in the remaining of the paper. It is argued in \cite[Chapter 2]{OST16} that it is valid at high frequencies when the wavelength $\lambda$ of the acoustic disturbances is much smaller than the characteristic length $L$ of the ambient quantities: this is precisely the situation analyzed here. \emphes{We also note that the model (\ref{eq:conv-wave}) has recently been considered in \cite{BOR19} to derive a high-frequency transport model for the acoustic energy density in a forward-scattering regime whereby the waves propagate within a cone of small opening angle}.
 
We thus consider acoustic disturbances in an open domain $\domain\subseteq\smash{\Rset^3}$, where $\Rset^3$ stands for the usual three-dimensional Euclidean space, which is constituted by an unsteady inhomogeneous ambient flow at the particle celerity $\vref(\xg,t)$, density $\rref(\xg,t)$, and speed of sound $\cref(\xg,t)$, for $\xg\in\domain$ and \emphes{$t\in\Rset$}. The velocity potential of the acoustic disturbances is denoted by $\smash{\vpote}(\xg,t)\in\Rset$, where the subscript $\escale$ stands for its (small) spatial and temporal scales of variation with respect to the scales of variation of the ambient quantities. That is, $\escale\equiv\smash{\frac{\lambda}{L}}$ and $\escale\equiv\smash{\frac{1}{\wg T}}$ where $\lambda$ and $\wg$ are the typical wavelength and circular frequency of the acoustic waves, respectively.
\emphes{Initial conditions for~\eref{eq:conv-wave} can be modeled by}:
\begin{equation}\label{eq:CI}
\vpote(\xg,0)=\vpot_0(\xg;\escale)\,,\quad\partial_t\vpote(\xg,0)=\psi_0(\xg;\escale)\,.
\end{equation}
They are parameterized by the small parameter $\escale$, which quantifies the rate of change of $\xg\mapsto\smash{\vpot_0}(\xg)$ and $\xg\mapsto\smash{\psi_0}(\xg)$ with respect to the typical length scale of the ambient flow. Since high-frequency waves will be generated by an initial acoustic disturbance oscillating at a scale $\escale\ll 1$, the functions $\smash{\bnabla_\xg\vpot_0}$ and $\smash{\psi_0}$ shall be considered as strongly $\escale$-oscillating functions in the sense of G\'erard \emph{et al}.~\cite{GER97}. The plane waves $\smash{\vpot_0(\xg;\escale)}=\smash{\escale A(\xg)\iexp^{\ci\kg\cdot\xg/\escale}}$ and $\smash{\psi_0(\xg;\escale)}=\smash{B(\xg)\iexp^{\ci\kg\cdot\xg/\escale}}$ for a given wave vector $\kg\in\smash{\Rset^3}$ and $\smash{\ci=\sqrt{-1}}$, typically fulfill this condition. \emphes{Prior to these excitations there is no wave, $\smash{\vpote}(\xg,t)\equiv 0$ for $t\ll 0$, but the background medium is moving due to the ambient flow}.

\section{\emphes{Pseudo-differential calculus and Wigner measure in a high-frequency setting}}\label{sec:PDO}

The high-frequency limit $\escale\rightarrow 0$ in the previous setting shall be derived for quadratic observables of the velocity potential $\smash{\vpote}$ as emphasized in \cite{AKI12,BAL02,BAL05,BAY14,BOR19,BRA02,FAN01,GUO99,POW05,RYZ96}. More particularly, the Wigner measure of the solutions of \eref{eq:PDOwave} shall be considered \cite{AKI12,GER97,LIO93,MAR02,RYZ96}, introducing a \emph{spatio-temporal} Wigner transform of that equation and its high-frequency limit as $\escale\rightarrow 0$ as in \emph{e.g.} \cite{BAL02,BAL05}. This is because the spatial and temporal scales in the wave equation \eqref{eq:PDOwave} play symmetric roles, and their oscillations should be accounted for altogether. Therefore a larger phase space than the usual phase space in physical space has to be introduced. In this section we summarize theses concepts and outline the main (formal) rules of pseudo-differential calculus that we shall use throughout the paper to characterize the Wigner measure and the transport and radiative transfer equations it satisfies.

\subsection{Semi-classical operators and Wigner measure}

From now on let us introduce the space-time variable $\xt=(\xg,t)\in\domain\times\Rset$ and its dual variable $\kw=(\kg,\wg)\in\Rset^4$ in the wave vector-frequency Fourier domain. Let $\obsj$ be a smooth, compactly supported function of both the space-time variable $\xt$ and impulse variable $\kw$. For a scalar field $\fu\in\smash{L^2(\Rset^4)}$, the functional space of square integrable functions endowed with the scalar product $\smash{\scald{\fu,\fv}}=\smash{\int_{\Rset^4}\fu(\xt)\cjg{\fv}(\xt)\,\dd\xt}$ where $\smash{\cjg{\fv}}$ stands for complex conjugation, consider the (semiclassical) operator:
\begin{equation}
\obsj^\vartheta(\xt,\escale\Dx)\fu(\xt)=\frac{1}{(2\pi)^4}\int_{\Rset^4\times\Rset^4}\iexp^{\ci\kw\cdot(\xt-\yt)}\obsj((1-\vartheta)\xt+\vartheta\yt,\escale\kw)\fu(\yt)\,\dd\yt\dd\kw\,,
\end{equation}
for $\vartheta\in[0,1]$. This parameter defines the so-called quantization of the operator. The case $\vartheta=0$ corresponds to the standard quantization. It is simply denoted by $\smash{\obsj(\xt,\escale\Dx)}$ such that:
\begin{equation}\label{eq:PDO}
\obsj(\xt,\escale\Dx)\fu(\xt)=\frac{1}{(2\pi)^4}\int_{\Rset^4}\iexp^{\ci\kw\cdot\xt}\obsj(\xt,\escale\kw)\TF{\fu}(\kw)\,\dd\kw\,,
\end{equation}
where:
\begin{equation}\label{eq:FT}
\TF{\fu}(\kw)=\int_{\Rset^4}\iexp^{-\ci\kw\cdot\xt}\fu(\xt)\dd\xt
\end{equation}
stands for the Fourier transform of $\fu(\xt)$. The case $\smash{\vartheta=1/2}$ corresponds to the Weyl quantization, which is usually denoted by $\smash{\obsj^W(\xt,\escale\Dx)}$. Then for a sequence $\smash{\sequence{\fue}}$ uniformly bounded in $\smash{L^2(\Rset^4)}$, there exists a positive measure $\Wigner[\fue]$ such that, up to extracting a subsequence if need be \alert{(see \emph{e.g.} \cite[Theorem 5.2]{ZWO12})}:
\begin{equation}\label{eq:mes-semi-classique}
\lim_{\escale\rightarrow 0}\scald{\obsj^\vartheta(\xt,\escale\Dx)\fue,\fue}=\int_{\Rset^4\times\Rset^4}\obsj(\xt,\kw)\Wigner[\fue](\dd\xt,\dd\kw)\,,\quad\forall\obsj\,,
\end{equation}
independently of the quantization $\vartheta$. \alert{Here the notation of \emph{e.g.} \cite[p.~330]{GER97} is used for the limit $\Wigner$ (independent of $\escale$) of the family $\smash{\sequence{\fue}}$  (dependent of $\escale$) but clearly the right-hand side in \eref{eq:mes-semi-classique} above is independent of $\escale$}. $\Wigner[\fue]$ is the so-called Wigner measure of $\smash{\sequence{\fue}}$ because it can also be interpreted as the weak limit of its Wigner transform $\smash{\Wignere^\vartheta[\fue,\fue]:=\Wignere^\vartheta[\fue]}$. Indeed, if the latter is defined for temperate distributions $\fu,\fv$ by:
\begin{equation}\label{eq:TWigner}
\Wignere^\vartheta[\fu,\fv](\xt,\kw)=\frac{1}{(2\pi)^4}\int_{\Rset^4}\iexp^{\ci\kw\cdot\yt}\fu\left(\xt-\escale(1-\vartheta)\yt\right)\adj{\fv}\left(\xt+\escale\vartheta\yt\right)\,\dd\yt\,,
\end{equation}
where $\adj{\fv}$ stands for the conjugate \emphes{(and transpose if it is a vector or a matrix)} of $\fv$, then one has the trace formula \cite{GER97}:
\begin{equation}\label{eq:trace}
\scald{\obsj^\vartheta(\xt,\escale\Dx)\fu,\fv}=\int_{\Rset^4\times\Rset^4}\obsj(\xt,\kw)\Wignere^\vartheta[\fu,\fv](\dd\xt,\dd\kw)\,.
\end{equation}
Thus $\smash{\Wigner[\vpote]}$ describes the limit energy of the sequence $\smash{\sequence{\vpote}}$ in the phase space $\smash{\Rset^4_\xt\times\Rset^4_\kw}$. The observable $\obsj(\xt,\kw)$ is used to select any quadratic observable or quantity of interest associated to this energy: the kinetic energy, or the free energy, or the power flow, \emph{etc}. \alert{This connection of Wigner measures with energetic quantities has already been outlined in \cite[Section~2.3]{BAY14} starting with the example of a simple oscillating function}. For example, the high-frequency "strain energy" $\smash{{\mathcal V}_\escale(t)}:=\smash{\demi\int_\domain\rref(\frac{1}{\cref}\dconv{\vpote})^2\,\dd\xg}$ in $\domain$ may be estimated from:
\begin{equation}\label{eq:strain-energy}
\lim_{\escale\rightarrow 0}{\mathcal V}_\escale(t)=\demi\int_{\domain\times\Rset^4}\rref(\xg,t)\Wigner\left[\frac{1}{\cref}\dconv{\vpote}(\cdot,t)\right](\dd\xg,\dd\kw)\,,
\end{equation}
up to some possible boundary effects on $\partial\domain$. Similarly, the "kinetic energy" $\smash{{\mathcal T}_\escale(t)}:=\smash{\demi\int_\domain\rref|\bnabla_\xg\vpote|^2\,\dd\xg}$ is estimated by:
\begin{equation}\label{eq:kinetic-energy}
\lim_{\escale\rightarrow 0}{\mathcal T}_\escale(t)=\demi\int_{\domain\times\Rset^4}\rref(\xg,t)\Wigner[\bnabla_\xg\vpote(\cdot,t)](\dd\xg,\dd\kw)\,.
\end{equation}
The acoustical energy density $\smash{{\mathcal E}_\escale}:=\smash{{\mathcal V}_\escale}+\smash{{\mathcal T}_\escale}$ that would be perceived by an observer moving with the ambient flow does not solve a closed-form equation in the high-frequency limit $\escale\rightarrow 0$. However the Wigner measure, which provides a decomposition of these quantities in phase space, does so as explained in the subsequent derivations. This is another reason why we shall now focus on such limit measure rather than $\smash{\vpote}$ directly or quadratic quantities of $\smash{\vpote}$.

\subsection{Some formal rules of pseudo-differential calculus}\label{sec:comp-adj}

In \cite{BAL05,BAY14} various formal rules of pseudo-differential calculus were given for quantities like $\smash{\Wignere^{1/2}[\obsj(\xg,\escale\Dxx)\fue,\fve]}$, mixing the quantization chosen by these authors for the Wigner transform (\ref{eq:TWigner}) ($\vartheta=1/2$) and the standard quantization of the semiclassical operator $\obsj(\xg,\escale\Dxx)$ ($\vartheta=0$). Here we revisit these formulas with the same quantization for both the Wigner transform and pseudo-differential operators, which seems more natural in view of the trace formula (\ref{eq:trace}). This will also significantly simplify the calculations below, leaving of course the final results unchanged since in the limit $\escale\rightarrow 0$ the Wigner measure is independent of the quantization; see \eref{eq:mes-semi-classique}. At last, we consider pseudo-differential operators in the space-time domain and its Fourier counterpart since, again, time and space play symmetric roles in \eref{eq:conv-wave}.

Let $\obsj(\xt,\kw)$ be a smooth function defined on $\smash{\Rset^4_\xt\times\Rset_\kw^4}$, and recall the notation of \eref{eq:PDO} for the operator $\smash{\obsj(\xt,\escale\Dx)}$ in the standard quantization $\vartheta=0$. Owing to \cite[Th. 2.7.4]{MAR02} for the composition of pseudo-differential operators, and \cite[Rem. 2.5.7]{MAR02} for the adjoint of a pseudo-differential operator, it can be shown that formally (omitting again the $\vartheta=0$ superscript for both $\smash{\Wignere}$ and $\smash{\obsj(\xt,\escale\Dx)}$):
\begin{multline}\label{reg1-cal}  
\Wignere[\obsj(\xt,\escale\Dx )\fue,\fve]= \obsj\left(\xt,\kw\right) \Wignere[\fue,\fve]-\frac{\escale}{\ci}\bnabla_\xt\obsj\cdot\bnabla_\kw\Wignere[\fue,\fve] \\
-\frac{\escale}{\ci}\left(\bnabla_\xt\cdot\bnabla_\kw\obsj\right)\Wignere[\fue,\fve]+\Go(\escale^2)\,, 
\end{multline}
and:
\begin{multline}\label{reg2-cal}  
\begin{split} 
\Wignere[\fue,\obsj( \xt, \escale \Dx) \fve ] &=\adj{\obsj}(\xt,\kw-\escale \Dx)\Wignere[\fue,\fve] \\
&= \adj{\obsj}(\xt,\kw)\Wignere[\fue,\fve]-\frac{\escale}{\ci}\bnabla_\kw\adj{\obsj}\cdot\bnabla_\xt\Wignere[\fue ,\fve] + \Go(\escale^{2})\,.
\end{split}
\end{multline}
Here $\bnabla_\xt A\cdot\bnabla_\kw B:= \bnabla_{\xg}A\cdot \bnabla_{\kg}B+\partial_t A\partial_\wg B$, and the differential operator $\smash{\Dx}$ within the observable $\obsj$ acts on $\smash{\Wignere[\fue,\fve]}$ so that, for instance, $\smash{\adj{\obsj}(\xt,\kw -\escale\Dx)\Wignere[\fue,\fve]}(\xt,\kw)$ should be interpreted as the inverse Fourier transform of $\smash{\adj{\obsj}(\xt,\kw - \escale\pw)}\smash{\TF{\Wigner}_\escale[\fue,\fve]}(\pw,\kw)$, where $\smash{\TF{\Wigner}_\escale[\fue,\fve](\pw,\kw)}$ is the Fourier transform \eqref{eq:FT} of $\smash{\Wigner_\escale[\fue,\fve](\xt,\kw)}$ with respect to the space-time variable $\xt$. For the chosen quantization, the Wigner transform is defined by:
\begin{equation}\label{eq:TWigner-xt}
\Wignere[\fue,\fve](\xt,\kw):=\frac{1}{(2\pi)^4}\int_{\Rset^4}\iexp^{\ci\kw\cdot\yt}\fue(\xt-\escale\yt) \adj{\fv}_\escale(\xt)\,\dd\yt\,.
\end{equation}
Note that if it is applied to $\fue$ and $\fve\equiv\fue$, $\Wignere[\fue,\fue]$ will be denoted by $\Wignere[\fue]$ as implicitly done in~\eref{eq:mes-semi-classique}.

The proof of these results is diverted to the \aref{apx:proof1}. If $\smash{\obsj(\xt,\escale\Dx)}$ is in addition formally self-adjoint the situation gets simpler and we have:
\begin{equation}\label{reg3-cal} 
\Wignere[\obsj(\xt,\escale\Dx )\fue,\fve]-\Wignere[\fue,\obsj( \xt, \escale \Dx) \fve ]=\frac{\escale}{\ci}\{\obsj,\Wignere[\fue,\fve]\}+\Go(\escale^2)\,,
\end{equation}
where $\{A,B\}:= \bnabla _{\kw}A\cdot \bnabla  _{\xt}B-\bnabla _{\xt}A\cdot \bnabla _{\kw}B$ stands for the usual Poisson bracket. This is also shown in the \aref{apx:proof1}. It should be noted at this stage that the second equality (\ref{reg2-cal}) above holds true if the Wigner transform $\smash{\Wignere[\fue,\fve]}$ does not depend on $\xt/\escale$. In \sref{sec:RTE} we will consider the situation where it also depends on this fast-scale variable and this rule has to be modified accordingly; this is done in \sref{sec:pse-dif-cal} below.

\subsection{Rules of pseudo-differential calculus with oscillating coefficients} \label{sec:pse-dif-cal}

Let $\yt\mapsto f(\yt)$ be a smooth real-valued function. Then, we have that: 
\begin{equation}\label{reg4-cal}
\begin{split}
\Wignere\left[f\left(\xte\right) \fue,\fve\right] &=  
 \frac{1}{(2\pi)^4} \int_{\Rset^4}  \iexp^{\ci\xte\cdot\pw} \TF{f} (\pw)  \Wignere[ \fue,\fve ] \left(\xt,\kw-\pw\right) \dd\pw \,,   \\  
\Wignere\left[\fue,f\left(\xte\right)\fve \right] &=f\left(\xte\right)\Wignere\left[\fue,\fve \right] \,,
\end{split}
\end{equation}
where $ \TF{f}$ is the Fourier transform \eqref{eq:FT} of $f$. Applying the above formula for highly oscillatory fluctuations $\xt/\escale$ with a real-valued observable $\smash{f(\yt)\obsj(\xt,\escale\Dx)}$ yields:
\begin{multline}\label{reg5-cal}
\Wignere\left[f\left(\xte\right)\obsj(\xt,\escale\Dx)\fue,\fve\right] \\
=\int\frac{\iexp^{\ci\xte\cdot\pw}\dd\pw}{(2\pi)^4}\TF{f}(\pw)\obsj(\xt,\kw-\pw)\Wignere[\fue,\fve]\left(\xt,\kw-\pw\right)+\Go(\escale)\,, 
\end{multline}
and:
\begin{equation}\label{reg5adj-cal}
\Wignere\left[\fue,f\left(\xte\right)\obsj(\xt,\escale\Dx)\fve\right] = f\left(\xte\right)\adj{\obsj}(\xt,\kw-\escale\Dx) \Wignere\left[\fue,\fve\right] \,,
\end{equation}
in view of \eref{reg1-cal} and \eref{reg2-cal} for example. These formulas will be used in the subsequent derivation of the evolution properties of the Wigner measure accounting for randomly perturbed ambient quantities. Their proofs are given in the \aref{apx:proof2}.

\section{Wigner measure of high-frequency acoustic waves in an unsteady inhomogeneous flow}\label{sec:Transport}

In this section we show how to obtain explicitly the Wigner measure~\eqref{eq:mes-semi-classique} of the high-frequency solutions of the convected wave equation~\eqref{eq:conv-wave} in the setting outlined in the foregoing section. The objectives are also to outline its main properties for a slowly varying ambient medium, as well as the (formal) mathematical tools used for its derivation. Both will prove useful in the subsequent~\sref{sec:RTE} where acoustic waves in a rapidly varying random ambient medium with correlation lengths comparable to the small wavelength $\escale$ are considered. The analysis presented here is derived from~\cite{GER97}, where first-order hyperbolic systems with constant and slowly varying coefficients are addressed, and~\cite{AKI12}, where arbitrary order hyperbolic systems with slowly varying coefficients are addressed. The dispersion properties of the acoustic Wigner measure are derived in~\sref{sec:dispersion}, and its evolution properties are derived in~\sref{sec:transport}. Here it is shown that it satisfies a Liouville transport equation, which states that the energy density in phase space is transported with a celerity corresponding to the convected group velocity. \emphes{We use the formal mathematical tools introduced in \sref{sec:comp-adj} above in order to compute the Wigner transform and its limit for high-frequency acoustic waves}. Now to hopefully clarify the subsequent derivations, we start by reformulating the convected wave equation (\ref{eq:conv-wave}) in a form that is adapted to the analysis developed in the remaining of the paper. 

\subsection{Acoustic wave equation as a semiclassical operator}\label{sec:PDO-wave-eq}

Here \eref{eq:conv-wave} is written in a more convenient form for the derivation of the high-frequency regime $\escale\ll 1$. We shall first consider a slowly fluctuating ambiant flow characterized by ambient quantities $\rref,\vref,\pref$ and $\cref$ which are independent of the small parameter $\escale$. Premultiplying it by $(\ci\escale)^2$, the acoustic wave equation (\ref{eq:conv-wave}) reads:
\begin{equation}\label{eq:PDOwave}
\vL_\escale(\xt,\escale\Dxt)\vpote=\bzero\,,\quad\xt=(\xg,t)\in\domain\times\Rset\,,
\end{equation}
where $\smash{\Dxt}=\smash{(\Dxx,\Dt)}$ for $\smash{\Dxx=\frac{1}{\ci}\bnabla_\xg}$, $\smash{\iD_t=\frac{1}{\ci}\partial_t}$, and the operator $\vL_\escale$ is:
\begin{equation}\label{eq:PDO-L}
\vL_\escale=\vL_0+\frac{\escale}{\ci}\vL_\indL\,, \\
\end{equation}
with:
\begin{equation}\label{eq:PDOperators}
\begin{split}
\vL_0(\xt,\kw;\cref,\vref) &=\demi\apar(\xt)\left(\vC^2(\xt,\kw)-\cref^2(\xt)\norm{\kg}^2\right)\,,\\
\vL_\indL(\xt,\kw;\cref,\vref) &=\demi\bnabla_\kw\cdot\bnabla_\xt\vL_0(\xt,\kw;\cref,\vref)\,, \\
\vC(\xt,\kw) &=\wg+\vref(\xt)\cdot\kg\,,
\end{split}
\end{equation}
and $\apar=\smash{\rref/\cref^2}>0$. \emphes{The impulse variable is $\kw=\smash{(\kg,\wg)\in\Rset^4}$ and its norm is $\norm{\kw}^2=\norm{\kg}^2+\wg^2$. We emphasizes that $\vL_0$ and $\vL_2$ depends on $\smash{\cref}$ and $\smash{\vref}$ which will both be considered as random fields in \sref{sec:RTE}}. \emphes{The expression of $\smash{\vL_\indL}$ stems from the formal self-adjointness of $\smash{\vL_\escale}$, which is shown in \aref{apx:proof-transport}}. Here and throughout we use the standard quantization $\vartheta=0$ and omit this superscript in all subsequent developments. It should be observed that $\smash{\vL_\indL(\xt,\escale\Dxt;\cref,\vref)}$ is a first-order partial differential operator in space and time, and that $\smash{\vL_\indL\equiv 0}$ in a steady, homogeneous medium where $\smash{\cref}$ and $\smash{\vref}$ are constant. In deriving \eref{eq:PDOwave} from \eref{eq:conv-wave} we have used \eref{eq:ambient-cons} and the fact that:
\begin{displaymath}
\rref\dconv{f}=\frac{\partial}{\partial t}(\rref f)+\bnabla_\xg\cdot(\rref f\vref)
\end{displaymath}
for any function $f$. Also the $1/2$ factor in $\smash{\vL_0}$ will be clarified below.

\subsection{Dispersion properties}\label{sec:dispersion}

The foregoing pseudo-differential calculus and spatio-temporal Wigner transform are now used for the wave equation (\ref{eq:PDOwave}). Computing the space-time Wigner transform~\eqref{eq:TWigner-xt} of $\vpote$ from \eref{eq:PDOwave}, yields:
\begin{equation*}
\Wignere\left[\vL_\escale(\xt,\escale\Dx_{\xt})\vpote,\vpote\right] = 0\,.
\end{equation*}
However, invoking the rule of calculus (\ref{reg1-cal}) we get:
\begin{multline}\label{eq:step0}
\Wignere[\vL_\escale(\xt,\escale\Dx_{\xt})\vpote,\vpote] = \left(\vL_0+\frac{\escale}{\ci}\vL_\indL\right)\Wignere[\vpote] - \frac{\escale}{\ci}\bnabla_\xt\vL_0\cdot\bnabla_\kw\Wignere[\fue]  \\
- \frac{\escale}{\ci}(\bnabla_\xt \cdot \bnabla_\kw\vL_0)\Wignere[\fue]+\Go(\escale^2)\,.
\end{multline}
Considering the leading-order term, one obtains as $\escale\rightarrow 0$ \alert{(see also for instance \cite[Theorem 5.3]{ZWO12})}:
\begin{equation}\label{eq:step01}
\vL_0(\xt,\kw)\Wigner[\fue](\xt,\kw)=0
\end{equation}
for the Wigner measure $\Wigner[\vpote]$ \alert{(independent of $\escale$)} of the sequence $\smash{\sequence{\vpote}}$ \alert{(dependent of $\escale$)} given by~\eref{eq:mes-semi-classique} in phase space $(\xt,\kw)\in\varXset=\domain\times\Rset_t\times\Rset^4_\kw\setminus\{\kg=\bzero\}$. \alert{Again, the notation of \cite[p. 330]{GER97} is used and the left-hand side in \eref{eq:step01} above is clearly independent of $\escale$}. Thus $\supp\smash{\Wigner[\vpote]}\subset\smash{\{(\xt,\kw)\in\varXset;\,\vL_0(\xt,\kw)=0\}}$, so the Wigner measure $\Wigner[\vpote]$ of the $\escale$-oscillating wave field $\smash{\sequence{\vpote}}$ is localized onto the different energy paths of the propagation operator with principal symbol $\smash{\vL_0(\xt,\kw)}$. These paths are determined by the equation $\smash{\vC(\xt,\kw)}=\pm\smash{\cref(\xt)\norm{\kg}}$ in phase space. They correspond to the rays for the ambient flow arising in classical Hamiltonian dynamics, as shown in the subsequent section. Hence $\Wigner[\vpote]$ may be written:
\begin{equation}\label{eq:Wignerpm}
\Wigner[\vpote](\xt,\kw) = \specij_-(\xt,\kg)\otimes\dir\left(\wg-\eigl_-(\xt,\kg)\right)+ \specij_+(\xt,\kg)\otimes\dir\left(\wg-\eigl_+(\xt,\kg)\right)\,,
\end{equation}
$\kg\neq\bzero$, where the $\specij_\pm$'s are the so-called specific intensities for high-frequency acoustic waves in an inhomogeneous ambient flow, which are independent of $\wg$, and:
\begin{equation}\label{eq:Doppler-freq}
\eigl_+(\xt,\kg)=-\vref(\xt)\cdot\kg-\cref(\xt)\norm{\kg}\,,\quad\eigl_-(\xt,\kg)=-\vref(\xt)\cdot\kg+\cref(\xt)\norm{\kg}
\end{equation}
are the Doppler frequencies. Note that the specific intensities are positive since the Wigner measure is positive. It is assumed in the remaining of the paper that these measures do not load the set $\{\kg=\bzero\}$.

\subsection{Evolution properties}\label{sec:transport}

On the other hand, we also have:
\begin{equation*}
\Wignere\left[\vpote,\vL_\escale(\xt,\escale\Dx_\xt)\vpote\right] = 0\,.
\end{equation*}
But $\smash{\vL_\escale}$ is formally self-adjoint, therefore using the rule (\ref{reg3-cal}) we end up with the simple Wigner equation:
\begin{equation}\label{eq:Wigner-transport}
0=\frac{\ci}{\escale}\left(\Wignere\left[\vL_\escale(\xt,\escale\Dx_\xt)\vpote,\vpote\right]-\Wignere\left[\vpote,\vL_\escale(\xt,\escale\Dx_\xt)\vpote\right]\right)=\{\vL_0,\Wignere[\vpote]\}+\Go(\escale)\,.
\end{equation}
Passing to the limit $\escale\rightarrow 0$ yields the transport equation in space-time phase space \alert{(see also for instance \cite[Theorem 5.4]{ZWO12})}:
\begin{equation}\label{eq:Wigner-equationb}
\{\vL_0,\Wigner[\vpote]\} = 0
\end{equation}
for the Wigner measure $\Wigner[\vpote]$ of the velocity quasi-potential $(\vpote)$. \alert{Here again with the notation of \cite[p. 330]{GER97}, the left-hand side above is clearly independent of $\escale$}. Introducing the following system of Hamiltonian equations:
\begin{equation}\label{eq:syst-Hamiltonian}
\begin{split}
\displaystyle\frac{\id\xt}{\id l} &=\bnabla_\kw\vL_0(\xt(l),\kw(l))\,, \\
\displaystyle\frac{\id\kw}{\id l} &=-\bnabla_\xt\vL_0(\xt(l),\kw(l))\,,
\end{split}
\end{equation}
with initial conditions satisfying $\vL_0(\xt(0),\kw(0))=0$, then its solutions $l\mapsto\smash{(\xt(l),\kw(l))}$ are the null bicharacteristics such that $\smash{\vL_0(\xt(l),\kw(l))}$ remains constant (and null) along them. Indeed, one observes by a straightforward application of the chain rule that \emphes{$\smash{\frac{\id}{\id l}\vL_0(\xt(l),\kw(l))\equiv0}$}. Thus the rays supporting the Wigner measure $\smash{\Wigner[\vpote]}$ may be constructed by solving the ordinary differential equations~(\ref{eq:syst-Hamiltonian}) (provided that the usual conditions for the local existence, uniqueness and smoothness of its solutions with respect to the initial conditions are fulfilled, see \emph{e.g.}~\cite{HAL80}. This issue is however much beyond the scope of this paper). The Wigner measure is kept constant on these \emphes{bicharacteristics} since by the chain rule again:
\begin{equation}\label{eq:Wigner-equationc}
\frac{\id\Wigner[\vpote]}{\id l}=\{\vL_0,\Wigner[\vpote]\}=0\,,
\end{equation}
\emphes{formally on the bicharacteristics}. In agreement with \cite{BRE69,CAN77,ENG74,GRO55,HAY68,KOR53,MIL21,PAP94,PIE90,THO65,UGI72}, the rays $l\mapsto(\xt(l),\kw(l))$ are determined by (this also clarifies the $\demi$ factor in \eref{eq:PDOperators}):
\begin{displaymath}
\begin{array}{ll}
\displaystyle\frac{\id\xg}{\id l}=\apar\vC\vref-\rref\kg\,, & \displaystyle\frac{\id\kg}{\id l}=\frac{\rref}{\cref}\norm{\kg}^2\bnabla_\xg\cref-\apar\vC\bnabla_\xg(\vref\cdot\kg)\,, \\
\displaystyle\frac{\id t}{\id l}=\apar\vC\,, & \displaystyle\frac{\id\wg}{\id l}=\frac{\rref}{\cref}\norm{\kg}^2\partial_t\cref-\apar\vC\partial_t\vref\cdot\kg \,,
\end{array}
\end{displaymath}
with \emphes{$\vC=\vC^+=-\cref\norm{\kg}$ if $\wg=\eigl_+$ or $\vC=\vC^-=+\cref\norm{\kg}$ if $\wg=\eigl_-$} on these paths \emphes{(in both cases $\vC\neq 0$)}. Alternatively, these ordinary differential equations may be written:
\begin{equation}\label{eq:rays}
\begin{array}{ll}
\displaystyle\frac{\id\xg}{\id t}=\vray^\pm\,, & \displaystyle\frac{\id\kg}{\id t}=-\bnabla_\xg(\vray^\pm\cdot\kg)\,, \\
\displaystyle\frac{\id l}{\id t}=\frac{1}{\apar\vC}\,, & \displaystyle\frac{\id\wg}{\id t}=-\partial_t(\vray^\pm\cdot\kg) \,,
\end{array}
\end{equation}
where $\smash{\vray^\pm}(\xt,\kg):=\smash{\vref(\xt)\pm\cref(\xt)\hkg}$ denotes the ray trajectory velocity, or group velocity \cite{BRE69,HAY68}, with the usual notation $\smash{\hkg=\kg/\norm{\kg}}$ for the unit vector in the direction of $\kg$. Note that $\smash{\eigl_\pm}(\xt,\kg)=-\smash{\vray^\pm}(\xt,\kg)\cdot\kg$ with these definitions. Here the time derivatives on the left sides are to be interpreted as time derivatives that would be seen when moving along a ray trajectory with the local and instantaneous group velocity $\smash{\vray^+}$ or $\smash{\vray^-}$. Relative to the ambient flow the ray moves in the direction of $\pm\smash{\kg}$ with the local and instantaneous speed of sound $\cref$. In this respect, the "total" time derivative of any quantity observed as moving on a ray trajectory in the forward ($+\smash{\hkg}$) or backward ($-\smash{\hkg}$) direction can be expressed in terms of its partial derivatives with respect to time and spatial coordinates as \cite{BRE69,PIE90}:
\begin{displaymath}
\frac{\id}{\id t}=\dconv{}\pm\cref\hkg\cdot\bnabla_\xg\,.
\end{displaymath} 

Now \eref{eq:rays} shows that if the ambient medium is frozen (\emph{i.e.} independent of $t$) then the wave frequency is unchanged, as expected. Alternatively, a time-dependent ambient medium is responsible for the \emph{spectral broadening}, or \emph{"haystacking"} effect of the acoustic spectrum around a tone frequency \cite{CAM78a,CAM78b,CAN76,GAR20,GUE85}. Spatial variations of the ambient flow velocity and sound speed are responsible for the \emph{phase shift} effect \cite{HOW73} which is manifested in the evolution of the wave vector $\kg$ along the paths. At last, introducing:
\begin{equation}\label{eq:defH}
\vH(\xt,\kw)=\demi\left(\vC^2(\xt,\kw)-\cref^2(\xt)\norm{\kg}^2\right)
\end{equation}
such that $\vL_0=\apar\vH$, the Liouville transport equation (\ref{eq:Wigner-equationc}) also reads (\emphes{formally on the bicharacteristics $t\mapsto(\xt(t),\kw(t))$):
\begin{equation}\label{eq:Wigner-equatione}
\frac{\id}{\id t}\left(\apar\vC\Wigner[\vpote]\right)=0\,.
\end{equation}
Alternatively, this transport equation is}:
\begin{equation}\label{eq:Wigner-equationd}
\{\vH,\apar\Wigner[\vpote]\}=0\,,
\end{equation}
as $\vH(\xt,\kw)=0$ on the support of $\smash{\Wigner[\vpote]}$. Indeed, \eref{eq:Wigner-equationb} holds in the sense of distributions that is $\smash{\int\{\vL_0,\obsj\}\dd\Wigner[\vpote]}=0$ for all smooth, compactly supported function $\obsj$ on $\varXset$. But $\smash{\{\vL_0,\obsj\}}=\apar\{\vH,\obsj\}+\vH\{\apar,\obsj\}$ and thus $\smash{\int\{\vL_0,\obsj\}\dd\Wigner[\vpote]}=\smash{\int\apar\{\vH,\obsj\}\dd\Wigner[\vpote]}=0$, $\forall\obsj$, which is the equality \eqref{eq:Wigner-equationd} above. \eref{eq:Wigner-equatione} is derived along the same lines from the first equality in \eref{eq:Wigner-equationc} with the change of variable of \eref{eq:rays}. In terms of the specific intensities $\smash{\specij_\pm}$ of \eref{eq:Wignerpm} the transport equation \eqref{eq:Wigner-equationd} is:
\begin{equation}\label{eq:Liouville-equation}
\partial_t\action_\pm+\bnabla_\kg(\vray^\pm\cdot\kg)\cdot\bnabla_\xg\action_\pm-\bnabla_\xg(\vray^\pm\cdot\kg)\cdot\bnabla_\kg\action_\pm=0\,,
\end{equation}
where it is easily verified that $\smash{\vray^\pm}(\xt,\kg)=\smash{\bnabla_\kg(\vray^\pm(\xt,\kg)\cdot\kg)}$, and:
\begin{equation}\label{eq:action}
\action_\pm(\xt,\kg)=\apar(\xt)\normu{\vC(\xt,\kg,\eigl_\pm(\xt,\kg))}\specij_\pm(\xt,\kg)=\frac{\rref(\xt)}{\cref(\xt)}\norm{\kg}\specij_\pm(\xt,\kg)
\end{equation}
are the forward and backward wave actions at the group velocities $\smash{\vray^+}=\smash{\vref+\cref\hkg}$ and $\smash{\vray^-}=\smash{\vref-\cref\hkg}$, respectively. Note that the ambient density $\rref$ does not influence the transport process in the proposed model since it only appears in the definition of $\smash{\action_\pm}$. \eref{eq:Liouville-equation} (or \eref{eq:Wigner-equationd}) is a generalization of \cite[Eq. (5)]{HAY68} accounting for angularly resolved wave actions in phase space and describes their conservation along the rays; see also \emph{e.g.} \cite[Eq. (1.9)]{BRE69}.

\section{Radiative transfer equation}\label{sec:RTE}

We now turn to the weak coupling regime of high-frequency acoustic waves in a random ambient flow. This regime denotes the situation whereby (i) propagation distances are large compared to the wavelength $\escale$, and (ii) random perturbations of the ambient quantities are weak and vary at the same scales as the wavelength (meaning that their correlation lengths scale as $\escale$). The subsequent analysis is derived from~\cite{BAL05} where acoustic waves in a quiescent medium are considered, and~\cite{BRA02} where general first-order anti-selfadjoint systems (possibly depending on time) are considered. The main result of this section is the radiative transfer equation (\ref{eq:RTE}) accounting for the \emph{multiple scattering} effect on acoustic waves in an ambient flow. Radiative transfer equations are linear Boltzmann equations which describe the kinetics of particles in a lattice of randomly distributed inclusions, for example. Thus high-frequency wave propagation phenomena may be very well understood in terms of a gas kinetics analogy. It involves collisional processes characterized in terms of differential and total scattering cross-sections, of which expressions are precisely given by \eref{eq:dscat} and \eref{eq:tscat}, respectively, for the present case. The different steps for this derivation are the following. The mathematical form chosen for modeling inhomogeneities is first given in the next \sref{sec:C1}. The random perturbations of the ambient quantities considered in the previous part are assumed to vary at the fast scale $\xg/\escale$ as opposed to the slow scale of variation $\xg$ of the latter. \emphes{Therefore one has to use (i) the rules of pseudo-differential calculus outlined in \sref{sec:pse-dif-cal} to account for both scales and generalize the rules invoked in the previous section; and (ii) a two-scale expansion of the Wigner transform of the waves in this situation. This is done in \sref{sec:random-Wigner-equation} and \sref{sec:multiple-scale-expansion}, respectively}. A major consequence of this separation of scales and of the scaling of the amplitudes of the random inhomogeneities in \sref{sec:C1} is that the fast scale does not modify the spectral (dispersion) properties of the Wigner measure already derived in \sref{sec:dispersion}. \sref{dis-rel-sec} shows why this property holds. However, the fast scale modifies the next-order contribution to the Wigner transform and consequently the evolution properties of the Wigner measure. The contribution of the fast scale of variations of the random perturbations of the ambient quantities to the two-scale expansion of the Wigner transform is given explicitly in \sref{fir-ord-cor-sec}. This correction actually gives rise to the collision operator characterizing the multiple scattering process of high-frequency waves on the random inhomogeneities. It is therefore responsible for the modification of the transport equation of \sref{sec:transport} for the bare ambient flow into a radiative transfer equation for the randomly perturbed ambient flow. The final \sref{second-corre-sec} outlines how this modification arises.

\subsection{Perturbations of the ambient flow}\label{sec:C1}
 
In the setting invoked above it is assumed that the ambient flow velocity and speed of sound now read:
\begin{equation}\label{ela-ten-fluc}
\begin{split}
\frac{1}{\cref^2(\xt)} &=\frac{1}{\Cref^2(\xt)}\left[1+\sqrt{\escale}\cper_1\left(\xte\right)\right] \,,\\
\vref(\xt) &=\Vref(\xt)+\sqrt{\escale}\Vper\left(\xte\right) \,.\\
\end{split}
\end{equation}
Here $\smash{\Vref}$ is the part of the ambient velocity varying at the slow scales $L$ and $T$ (outer scale of turbulence), and \alert{$\smash{\Vper}$} is its fluctuation with amplitude $\sqrt{\escale}$. \alert{Likewise, $\smash{\Cref}$ is the ambient speed of sound and $\smash{\cper_1}$ is the fluctuations of its squared inverse with amplitude $\sqrt{\escale}$}. We typically think of $\smash{\Vref}$ as the flow component on the energy-containing integral scale, and $\sqrt{\escale}\Vper$ as the flow component on the inertial subrange. In this approach the typical acoustic wavelength $\lambda$ lies in the inertial subrange $\ell\ll\lambda\ll L$ where $\ell$ is the inner scale of turbulence, or dissipation (Kolmogorov) length, and $\escale\equiv\frac{\lambda}{L}\ll 1$. Note that the proposed simple model of turbulence is the one retained in \cite{FAN01}. \emphes{Our scaling is also different from the one considered in \cite{BOR19}, where a forward-scattering regime of propagation is emphasized for a constant mean flow velocity $\smash{\Vref}$ independent of $\xt$}. Of course such a separation of scales may not be feasible for real turbulence.

The fluctuations $\smash{\vper}=\smash{(\cper_1,\Vper)}$ are modeled by a vector-valued, second-order stochastic field $\smash{\{\vper(\yt)\,;\,\yt\in\Rset^4\}}$, which has mean zero and is mean square homogeneous (stationary). The latter property means that the cross-correlations of the perturbations at two different locations and times $\yt$ and $\yt'$ depend on $\yt'-\yt$ solely; it is the model actually retained in \emph{e.g.} \cite{FAN01,HOW73,OST16}. In \cite{HOW73} the author considered either perturbations of the ambient velocity associated with the presence of sound waves, or perturbations of the sound speed associated with random variations of the ambient temperature. If the cross-correlations depend on $|\yt-\yt'|$ the medium is said to be statistically isotropic. At last, the inhomogeneities are small as expressed by their $\smash{\Go(\escale^\demi)}$ amplitude. This size is the unique scaling which allows them to significantly modify the energy spreading in the transport regime at long propagation distances; see \emph{e.g.} \cite{RYZ96,BAL05}. Larger fluctuations could lead to localization of the waves, a situation beyond the scope of kinetic models. The model of correlation of the fluctuation velocity is given as:
\begin{equation}\label{def-aver}
\esp{\smash{\TF{\vg}_1(\kw)\otimes\TF{\vg}_1(\pw)}}:= (2\pi)^8\dir(\kw+\pw) \TF{\coro}(\kw)\,,
\end{equation}
where the $4\times 4$ correlation tensor is $\smash{\coro(\yt'-\yt)} := \esp{\smash{\vper(\yt)\otimes\vper(\yt')}}$ and:
\begin{equation*}
\TF{\coro}(\kw) := \frac{1}{(2\pi)^4}\int_{\Rset^4}\iexp^{\ci\yt\cdot\kw}\coro(\yt)\dd\yt=\begin{bmatrix} \TF{\coroij}_c(\kw) & \adj{\TF{\coro}}_{cv}(\kw) \\ \TF{\coro}_{cv}(\kw) & \TF{\coro}_v(\kw) \end{bmatrix}\,.
\end{equation*}
In the above $\esp{\cdot}$ stands for the mathematical expectation (average), $\smash{\TF{\coroij}_c(\kw)}$ is the power spectral density of the perturbations of the speed of sound, $\smash{\TF{\coro}_v(\kw)}$ is the power spectral density matrix of the perturbations of the particle velocity, and $\smash{\TF{\coro}_{cv}(\kw)}$ is the cross-spectral density vector of the perturbations of the speed of sound and particle velocity. We stress that the spectral density matrix $\vp\mapsto\smash{\TF{\coro}(\kw)}$ is such that $\smash{\TF{\coro}(-\kw)}=\smash{\TF{\coro}(\kw)^\itr}$, where $\smash{{\bf A}^\itr}$ stands for the transpose of matrix ${\bf A}$. \emphes{As in \cite{RYZ96} we assume that $\smash{\TF{\coro}}$ is real, which implies that both $\smash{\TF{\coro}}$ and $\smash{\coro}$ are even functions: $\smash{\TF{\coro}}(-\kw)=\smash{\TF{\coro}}(\kw)$ and $\coro(-\yt)=\coro(\yt)$}. Usually it is further assumed that $\smash{\TF{\coro}_{cv}(\kw)}=\bzero$ (the perturbations of the speed of sound and of the particle velocity are uncorrelated), which is true if these perturbations are statistically isotropic and incompressible, $\bnabla_\xg\cdot\Vper=0$ \cite{OST16,WEN71}. If the latter properties hold, the power spectral density matrix of the perturbations of the particle velocity reads $\smash{\TF{\coro}_v(\kw)}=\smash{\TF{\mathcal{R}}(\norm{\kg},\wg)(\II-\hat\kg\otimes\hat\kg)}$, where $\smash{\TF{\mathcal{R}}(\norm{\kg},\wg)}$ is a scalar function and $\kw=(\kg,\wg)$ \emph{i.e.} wave vector and (circular) frequency.

We note at this stage that a randomly perturbed ambient density may be accounted for as well in the subsequent developments. However it is observed, following the conclusions of \sref{sec:transport}, that the ambient density does not influence the evolution of the Wigner measure in our model. Thus we ignore that possibility in the remaining of the paper. The analysis could be carried on, though, along the same lines as in~\cite[Sect. 7]{BAL05}. \alert{Also the consideration of non-homogeneous (non stationary) perturbations--provided that relevant models of non homogeneous turbulent fluctuations are available-- is left to future works as it may yield very different propagation regimes}.

\subsection{Acoustic wave equation with randomly perturbed ambient quantities}\label{sec:random-Wigner-equation}

Having introduced the random fluctuations of the ambient flow velocity, we can write the acoustic wave equation (\ref{eq:conv-wave}) accounting for these inhomogeneities in a similar form as \eref{eq:PDOwave} as follows. Let us introduce two operators $\smash{\vL_1}$ and $\smash{\vL_{\indL 1}}$ arising from the random fluctuations $\smash{\vper}$ and defined by:
\begin{equation}\label{eq:PDO-pert}
\begin{split}
\vL_1\left(\xt,\xte,\escale\Dx_\xt\right) &=\demi\apar(\xt)\left[\vC(\xt,\escale\Dx_\xt)\Vper\left(\xte\right)\cdot(\escale\Dx_\xg)+\Vper\left(\xte\right)\cdot(\escale\Dx_\xg)\vC(\xt,\escale\Dx_\xt)\right. \\
&\quad\quad\quad\quad\left.+\vC(\xt,\escale\Dx_\xt)\cper_1\left(\xte\right)\vC(\xt,\escale\Dx_\xt)\right]\,, \\
\vL_{\indL 1}\left(\xt,\xte,\escale\Dx_\xt\right) &=\frac{\ci}{2}\apar(\xt)\left[\Vper\left(\xte\right)\cdot(\escale\Dx_\xg)\cper_1\left(\xte\right)\vC(\xt,\escale\Dx_\xt)+\left(\Vper\left(\xte\right)\cdot\escale\Dx_\xg\right)^2\right. \\
&\quad\quad\quad\quad\left.+\vC(\xt,\escale\Dx_\xt)(\cper_1\Vper)\left(\xte\right)\cdot(\escale\Dx_\xg)\right]\,,
\end{split}
\end{equation}
where $\apar(\xt)=\smash{\rref(\xt)/\Cref^2(\xt)}$ and $\vC(\xt,\kw)=\wg+\Vref(\xt)\cdot\kg$, as in \eref{eq:PDOperators}. The acoustic wave equation (\ref{eq:PDOwave}) is now $\smash{\mathcal{\vL}_\escale\vpote}=0$ with the operator $\smash{\mathcal{\vL}_\escale}$ defined by:
\begin{equation}\label{eq:PDO-L-RTE}
\begin{split}
\mathcal{\vL}_\escale &= \vL_\escale +\sqrt{\escale}\vL_1+\frac{\escale}{\ci}\vL_{\indL 1}+\Go(\escale^\frac{3}{2}) \\
&=\vL_0+\sqrt{\escale}\vL_1+\frac{\escale}{\ci}(\vL_\indL+\vL_{\indL 1})+\Go(\escale^\frac{3}{2})\,,
\end{split}
\end{equation}
where $\smash{\vL_\escale}$ is given by \eref{eq:PDO-L} with the unperturbed ambient quantities $\smash{\Cref}$ and $\smash{\Vref}$: \emphes{$\smash{\vL_\escale}(\xt,\kw)=\smash{\vL_0}(\xt,\kw;\Cref,\Vref)-\ci\escale\smash{\vL_\indL}(\xt,\kw;\Cref,\Vref)$}.
Then by applying the space-time Wigner transforms $\smash{\Wignere[\cdot,\vpote]}$ and $\smash{\Wignere[\vpote,\cdot]}$  to \eref{eq:PDOwave} with $\smash{\mathcal{\vL}_\escale}$ given by~\eref{eq:PDO-L-RTE}, we obtain respectively:
\begin{multline}\label{sys-wig-dr-ga}
\Wignere[\vL_\escale( \xt,\escale\Dx_\xt) \vpote,\vpote] + \sqrt{\escale} \Wignere\left[\vL_1\left(\xt,\xte,\escale \Dx_\xt\right)\vpote,\vpote\right] \\
+\frac{\escale}{\ci}\Wignere\left[\vL_{\indL 1}\left(\xt,\xte,\escale\Dx_\xt\right)\vpote,\vpote\right]=\Go(\escale^\frac{3}{2})\,,
\end{multline}
and:
\begin{multline}\label{sys-wig-dr-ga-adj}
\Wignere[\vpote,\vL_\escale( \xt,\escale\Dx_\xt) \vpote] + \sqrt{\escale} \Wignere\left[\vpote,\vL_1\left(\xt,\xte,\escale \Dx_\xt\right)\vpote\right] \\
-\frac{\escale}{\ci}\Wignere\left[\vpote,\vL_{\indL 1}\left( \xt,\xte,\escale\Dx_\xt\right) \vpote\right]=\Go(\escale^\frac{3}{2})\,.
\end{multline}
The operator $\smash{\vL_\escale}$ corresponding to the unperturbed wave equation \emphes{is formally self-adjoint because the unperturbed ambient quantities $\smash{\Cref}$ and $\smash{\Vref}$ satisfy Eq.~(\ref{eq:ambient-cons}$a$)} (see \aref{apx:proof-transport}); it is in addition independent of $\xt/\escale$. Therefore taking the difference of \eref{sys-wig-dr-ga} and \eref{sys-wig-dr-ga-adj} yields:
\begin{multline}\label{eq:Wigner-transfer}
\Wignere[\vL_\escale( \xt,\escale\Dx_\xt) \vpote,\vpote]-\Wignere[\vpote,\vL_\escale( \xt,\escale\Dx_\xt) \vpote] \\
+\sqrt{\escale} \Wignere\left[\vL_1\left(\xt,\xte,\escale \Dx_\xt\right)\vpote,\vpote\right]-\sqrt{\escale} \Wignere\left[\vpote,\vL_1\left(\xt,\xte,\escale \Dx_\xt\right)\vpote\right] \\
+\frac{\escale}{\ci}\Wignere\left[\vL_{\indL 1}\left(\xt,\xte,\escale\Dx_\xt\right)\vpote,\vpote\right]+\frac{\escale}{\ci}\Wignere\left[\vpote,\vL_{\indL 1}\left( \xt,\xte,\escale\Dx_\xt\right) \vpote\right]=\Go(\escale^\frac{3}{2})\,.
\end{multline}
\eref{eq:Wigner-transfer} for the case of a randomly inhomogeneous medium is the counterpart of the Wigner equation (\ref{eq:Wigner-transport}) for a slowly varying ambient flow. The difference stems from the terms involving $\smash{\vL_1}$ and $\smash{\vL_{\indL 1}}$, which must be carefully evaluated in an asymptotic analysis since they contain both scales $\xt$ and $\xt/\escale$. We may then make use of the rules of pseudo-differential calculus with oscillating coefficients described in \sref{sec:pse-dif-cal}. In addition, the Wigner transform $\smash{\Wignere[\vpote]}$ also depends on these two scales, and therefore its derivatives with respect to $\xt$ must be carefully evaluated as well. This issue is subsequently addressed in \sref{sec:multiple-scale-expansion} below. 

\subsection{Multiple scale expansion of the Wigner transform of the acoustic wave equation}\label{sec:multiple-scale-expansion}

Now we introduce a two-scale version of $\smash{\Wignere[\vpote]}$ as follows:
\begin{equation}\label{asym-expan}
\begin{split}
\Wignere[\vpote](\xt,\kw) &=\tilde{\Wigner}_\escale\left(\xt,\yt,\kw\right) \\
&=\Wigner_0(\xt,\kw) + \sqrt{\escale} \Wigner_{1}(\xt,\yt,\kw) + \escale\Wigner_{2}(\xt,\yt,\kw) + \po(\escale)\,.
\end{split}
\end{equation}
Consequently, the space-time gradient $\escale\Dx_\xt$ has to be rewritten $\smash{\escale\Dx_\xt+\Dx_\yt}$ to account for this two-scale expansion in \eref{eq:Wigner-transfer}. Considering the terms corresponding to the unperturbed wave equation in this latter equation, one first has, invoking the rules (\ref{reg1-cal})--(\ref{reg2-cal}):
\begin{multline*}
\Wignere[\vL_\escale( \xt,\escale\Dx_\xt) \vpote,\vpote]-\Wignere[\vpote,\vL_\escale( \xt,\escale\Dx_\xt) \vpote]= \left(\vL_0(\xt,\kw)-\adj{\vL}_0(\xt,\kw-\escale\Dx_\xt)\right)\tilde{\Wigner}_\escale \\
-\frac{\escale}{\ci}\bnabla_\kw\cdot(\bnabla_\xt\vL_0(\xt,\kw)\,\tilde{\Wigner}_\escale) + \frac{\escale}{\ci}\left(\vL_\indL(\xt,\kw)+\adj{\vL}_\indL(\xt,\kw-\escale\Dx_\xt)\right)\tilde{\Wigner}_\escale +\Go(\escale^2) \\
= \left(\vL_0(\xt,\kw)-\vL_0(\xt,\kw-\Dx_\yt)\right)\tilde{\Wigner}_\escale+\frac{\escale}{\ci}\bnabla_\kw\vL_0(\xt,\kw-\Dx_\yt)\cdot\bnabla_\xt\tilde{\Wigner}_\escale \\
-\frac{\escale}{\ci}\bnabla_\kw\cdot(\bnabla_\xt\vL_0(\xt,\kw)\,\tilde{\Wigner}_\escale)+ \frac{\escale}{\ci}\left(\vL_\indL(\xt,\kw)+\vL_\indL(\xt,\kw-\Dx_\yt)\right)\tilde{\Wigner}_\escale +\Go(\escale^2)\,.
\end{multline*}
Considering now the third and fourth terms in \eref{eq:Wigner-transfer}, one obtains using the rules (\ref{reg5-cal}) and (\ref{reg5adj-cal}) and the notation $\yt=\xt/\escale=(\yg,u)$:
\begin{multline}\label{reg5tot-cal}
\Wignere\left[\vL_1\left(\xt,\yt,\escale \Dx_\xt\right)\vpote,\vpote\right]-\Wignere\left[\vpote,\vL_1\left(\xt,\yt,\escale \Dx_\xt\right)\vpote\right]= \\
\frac{\apar(\xt)}{2}\int_{\Rset^4} \frac{\iexp^{\ci\yt\cdot\pw}\dd\pw}{(2\pi)^4}\acoeff(\xt,\kw,\pw,\kw-\pw)\tilde{\Wigner}_\escale\left(\xt,\yt,\kw-\pw\right) \\
-\frac{\apar(\xt)}{2}\Big[\vC(\xt,\kw-\Dx_\yt)\Vper(\yt)\cdot(\kg-\Dx_\yg)+\Vper(\yt)\cdot(\kg-\Dx_\yg)\vC(\xt,\kw-\Dx_\yt) \\
+\vC(\xt,\kw-\Dx_\yt)\cper_1(\yt)\vC(\xt,\kw-\Dx_\yt)\Big]\tilde{\Wigner}_\escale(\xt,\yt,\kw) + \Go(\escale)\,,
\end{multline}
where (remind \eref{eq:PDO-pert}):
\begin{equation}\label{eq:def-alpha}
\acoeff(\xt,\kw,\pw,\kw')=\vC(\xt,\kw+\kw')\TF{\Vg}_1(\pw)\cdot\kg'+\vC(\xt,\kw)\TF{\cper}_1(\pw)\vC(\xt,\kw')\,.
\end{equation}
In the above we have noted $\kw=(\kg,\wg)$, $\kw'=(\kg',\wg')$, and $\pw=(\vp,\upsilon)$ so that $\kw-\pw=(\kg-\vp,\wg-\upsilon)$. At last, one notices again from \eref{reg5-cal} and \eref{reg5adj-cal} that:
\begin{multline}\label{reg6-cal}
\Wignere\left[\vL_{\indL 1}\left(\xt,\yt,\escale \Dx_\xt\right)\vpote,\vpote\right]+\Wignere\left[\vpote,\vL_{\indL 1}\left(\xt,\yt,\escale \Dx_\xt\right)\vpote\right]= \\
\frac{\ci\apar(\xt)}{2}\int_{\Rset^8} \frac{\iexp^{\ci\yt\cdot(\pw+\pw')}\dd\pw\dd\pw'}{(2\pi)^8}\TF{\beta}(\xt,\kw,\pw,\pw')\tilde{\Wigner}_\escale\left(\xt,\yt,\kw-\pw-\pw'\right) \\
-\frac{\ci\apar(\xt)}{2}\Big[\Vper(\yt)\cdot(\kg-\Dx_\yg)\cper_1(\yt)\vC(\xt,\kw-\Dx_\yt)+\left(\Vper(\yt)\cdot(\kg-\Dx_\yg)\right)^2 \\
\quad\quad\quad\quad+\vC(\xt,\kw-\Dx_\yt)(\cper_1\Vper)(\yt)\cdot(\kg-\Dx_\yg)\Big]\tilde{\Wigner}_\escale(\xt,\yt,\kw) + \Go(\escale) \,,
\end{multline}
where $\pw'=(\vp',\upsilon')$ and:
\begin{displaymath}
\begin{split}
\TF{\beta}(\xt,\kw,\pw,\pw')&=\TF{\Vg}_1(\pw)\cdot(\kg-\pg)\left[\TF{\cper}_1(\pw')\vC(\xt,\kw-\pw-\pw')+\TF{\Vg}_1(\pw')\cdot(\kg-\pg-\pg')\right] \\
&\quad+\vC(\xt,\kw)\TF{\cper}_1(\pw)\TF{\Vg}_1(\pw')\cdot(\kg-\pg-\pg') \,.
\end{split}
\end{displaymath}

Thus the Wigner equation (\ref{eq:Wigner-transfer}) takes the form:
\begin{multline}\label{second-equ}
\left[\vL_0(\xt,\kw)-\vL_0(\xt,\kw-\Dx_\yt)\right]\tilde{\Wigner}_\escale+\sqrt{\escale}{\mathcal L}_1\tilde{\Wigner}_\escale \\
+\frac{\escale}{\ci}\left[\bnabla_\kw\vL_0(\xt,\kw-\Dx_\yt)\cdot\bnabla_\xt-\bnabla_\xt\vL_0(\xt,\kw)\cdot\bnabla_\kw-\bnabla_\kw\cdot\bnabla_\xt\vL_0(\xt,\kw)\right]\tilde{\Wigner}_\escale \\
+\frac{\escale}{\ci}\left[\vL_\indL(\xt,\kw)+\vL_\indL(\xt,\kw-\Dx_\yt)\right]\tilde{\Wigner}_\escale +  \frac{\escale}{\ci}{\mathcal L}_{21}\tilde{\Wigner}_\escale = \Go(\escale^\frac{3}{2})\,,
\end{multline}
where $\smash{{\mathcal L}_1}$ is given by the right-hand side of \eref{reg5tot-cal} and $\smash{{\mathcal L}_{21}}$ is given by the right-hand side of \eref{reg6-cal}. Now we equate like-powers of $\escale$ in \eref{second-equ} to obtain a sequence of three equations for the orders $\smash{\Go(\escale^0)}$, $\smash{\Go(\escale^\demi)}$ and $\smash{\Go(\escale)}$. This procedure follows~\cite[Sect.~7]{BAL05} for quiescent acoustics. Thus we can follow the analysis developed in~\sref{sec:Transport} to account for the influence of the random perturbations characterized by the operators $\smash{\vL_1}$ and $\smash{\vL_{\indL 1}}$ above. The $\smash{\Go(\escale^0)}$ terms yield the dispersion properties of $\smash{\Wigner_0}$ as in \sref{sec:dispersion}, while the $\smash{\Go(\escale)}$ terms yield its evolution properties as in~\sref{sec:transport}. The $\smash{\Go(\escale^\demi)}$ terms yield a linear relation between $\smash{\Wigner_1}$ and $\smash{\Wigner_0}$ that explicit the contribution of the random inhomogeneities on the evolution properties of the latter.

\subsection{Dispersion properties}\label{dis-rel-sec}

We start by establishing the connection between the temporal and spatial oscillations of the waves in the high-frequency limit, the so-called dispersion relation. It is given by the leading order terms $\smash{\Go(\escale^0)}$ in~\eref{sys-wig-dr-ga}. Since the symbol of the operator $\vL_0$ of \eref{eq:PDO-L-RTE} is identical with that of \eref{eq:step0}, we simply have $\vL_0(\xt,\kw)\Wigner_0(\xt,\kw)=0$ and thus, as with \eref{eq:Wignerpm}:
\begin{equation}\label{dis-rel-3}
\Wigner_0(\xt,\kw) = \specij_-(\xt,\kg)\otimes\dir\left(\wg-\eigl_-(\xt,\kg)\right)+ \specij_+(\xt,\kg)\otimes\dir\left(\wg-\eigl_+(\xt,\kg)\right)\,,
\end{equation}
where $\smash{\eigl_\pm}$ are given by \eref{eq:Doppler-freq} with $\cref$ and $\vref$ replaced by $\Cref$ and $\Vref$, respectively. We keep the same notations for the specific intensities $\smash{\specij_\pm}$ as in the non-random case for convenience.


\subsection{Half-order correction $\Go(\escale^\demi)$}\label{fir-ord-cor-sec}

By considering the $\smash{\Go(\escale^\demi)}$ terms in~\eref{second-equ} we can calculate $\smash{\TF{\Wigner}_{1}(\xt,\vp,\kw)}$, the Fourier transform of $ \smash{\Wigner_1}(\xt,\yt,\kw)$ with respect to $\yt$, in terms of $\smash{\Wigner_0}(\xt,\kw)$. This expression will be used in the sequel for the derivation of the evolution properties of $\smash{\Wigner_0}$. The $\smash{\Go(\escale^\demi)}$ terms are:
\begin{multline*} \label{cal-1-s-ins}
0=\left(\vL_0(\xt,\kw)-\vL_0(\xt,\kw-\Dx_\yt)\right)\Wigner_1(\xt,\yt,\kw) \\
+\frac{\apar(\xt)}{2}\int_{\Rset^4} \frac{\iexp^{\ci\yt\cdot\pw}\dd\pw}{(2\pi)^4}\acoeff(\xt,\kw,\pw,\kw-\pw)\Wigner_0\left(\xt,\kw-\pw\right) \\
-\frac{\apar(\xt)}{2}\Big[\left(\vC(\xt,\kw-\Dx_\yt)+\vC(\xt,\kw)\right)\Vper(\yt)\cdot\kg+\vC(\xt,\kw-\Dx_\yt)\cper_1(\yt)\vC(\xt,\kw)\Big]\Wigner_0(\xt,\kw)
\end{multline*}
since $\smash{\Wigner_0}$ is independent of $\yt$. Taking the Fourier transform with respect to $\yt$ yields:
\begin{equation}\label{coef-w1}
\TF{\Wigner}_1(\xt,\pw,\kw)=\demi\frac{\acoeff(\xt,\kw-\pw,\pw,\kw)\Wigner_0(\xt,\kw)-\acoeff(\xt,\kw,\pw,\kw-\pw)\Wigner_0(\xt,\kw-\pw)}{\vH(\xt,\kw)-\vH(\xt,\kw-\pw)-\ci\theta}\,.
\end{equation}   
Here $\theta$ is a regularization parameter to evade the case $\smash{\vH(\xt,\kw)}=\smash{\vH(\xt,\kw-\pw)}$ for the time being. It will be sent to $0$ at the end of the derivation.


\subsection{Evolution properties}  \label{second-corre-sec}

The evolution equation for $\smash{\Wigner_0}$ is finally obtained from the $\Go(\escale)$ terms in \eref{second-equ}. The latter are:
\begin{equation}\label{second-correc-1}
\begin{split}
0=& \frac{1}{\ci}\{\vL_0,\Wigner_0\}+\frac{1}{\ci}\left(2\vL_\indL(\xt,\kw)-\bnabla_\kw\cdot\bnabla_\xt\vL_0(\xt,\kw)\right)\Wigner_0(\xt,\kw) \\
&+\left(\vL_0(\xt,\kw)-\vL_0(\xt,\kw-\Dx_\yt)\right)\Wigner_2(\xt,\yt,\kw)  \\
&+\frac{\apar(\xt)}{2}\int_{\Rset^4} \frac{\iexp^{\ci\yt\cdot\pw}\dd\pw}{(2\pi)^4}\acoeff(\xt,\kw,\pw,\kw-\pw)\Wigner_1\left(\xt,\yt,\kw-\pw\right) \\
&-\frac{\apar(\xt)}{2}\Big[\vC(\xt,\kw-\Dx_\yt)\Vper(\yt)\cdot(\kg-\Dx_\yg)+\Vper(\yt)\cdot(\kg-\Dx_\yg)\vC(\xt,\kw-\Dx_\yt) \\
&\quad\quad\quad\quad+\vC(\xt,\kw-\Dx_\yt)\cper_1(\yt)\vC(\xt,\kw-\Dx_\yt)\Big]\Wigner_1(\xt,\yt,\kw) \\
&+\frac{\apar(\xt)}{2}\int_{\Rset^8} \frac{\iexp^{\ci\yt\cdot(\pw+\pw')}\dd\pw\dd\pw'}{(2\pi)^8}\TF{\beta}(\xt,\kw,\pw,\pw')\Wigner_0\left(\xt,\kw-\pw-\pw'\right) \\
&-\frac{\apar(\xt)}{2}\Big[\Vper(\yt)\cdot(\kg-\Dx_\yg)\cper_1(\yt)\vC(\xt,\kw)+\Vper(\yt)\cdot(\kg-\Dx_\yg)\Vper(\yt)\cdot\kg \\
&\quad\quad\quad\quad+\vC(\xt,\kw-\Dx_\yt)(\cper_1\Vper)(\yt)\cdot\kg\Big]\Wigner_0(\xt,\kw)\,.
\end{split}   
\end{equation}
The term in $\smash{\Wigner_2}$ vanishes once it is averaged ($\smash{\esp{\Wigner_2}}=0$ since $\smash{\esp{\smash{\tilde{\Wigner}_\escale}}}=\smash{\esp{\Wigner_0}}$ \emphes{by construction}). The sum of the last two terms in $\smash{\Wigner_0}$ vanishes as well once it is averaged in view of \eref{def-aver} and the mixing assumption invoked below, see \eref{eq:mixing}. Also $\smash{2\vL_\indL-\bnabla_\kw\cdot \bnabla_\xt\vL_0}=0$ \emphes{by the formal self-adjointness of $\smash{\vL_\escale}$, see \eref{eq:PDOperators}}, hence only the Poisson bracket with $\smash{\Wigner_0}$ remains. Thus the last step is to evaluate the integrals above when $\smash{\Wigner_1}$ is replaced by its expression (\ref{coef-w1}) as a function of $\smash{\Wigner_0}$. This closure, together with averaging in~\eref{second-correc-1}, gives rise to a collisional linear radiative transfer equation for the average $\esp{\smash{\Wigner_0}}$ of the Wigner measure associated to the waves velocity potential. The collision operator is shown to depend on the phase functions of the random ambient inhomogeneities, $\smash{\TF{\coro}(\pw)}$ in \eref{def-aver}. We obtain here a general form of the collisional kernel describing multiple scattering of waves in a random, unsteady ambient flow which is the main contribution of the paper. We detail in the next two subsections how it is derived.


\subsubsection{Averaging \eref{second-correc-1}}

We first consider the following integral term in \eref{second-correc-1}:
\begin{equation*}
I_1=\int_{\Rset^4} \frac{\iexp^{\ci\yt\cdot\pw}\dd\pw}{(2\pi)^4}\acoeff(\xt,\kw,\pw,\kw-\pw)\Wigner_1\left(\xt,\yt,\kw-\pw\right)\,.
\end{equation*}
Introducing the Fourier transform $\smash{\TF{\Wigner}_1}$ of $\smash{\Wigner_1}$ given by \eref{coef-w1} yields:
\begin{equation*}
\begin{split}
I_1 &=\int_{\Rset^8} \frac{\iexp^{\ci\yt\cdot(\pw+\pw')}\dd\pw\dd\pw'}{(2\pi)^8}\acoeff(\xt,\kw,\pw,\kw-\pw)\TF{\Wigner}_1(\xt,\pw',\kw-\pw) \\
&=\demi\int_{\Rset^8} \frac{\iexp^{\ci\yt\cdot(\pw+\pw')}\dd\pw\dd\pw'}{(2\pi)^8}\frac{\acoeff\left(\xt,\kw,\pw,\kw-\pw\right)}{\vH(\xt,\kw-\pw)-\vH(\xt,\kw-\pw-\pw')-\ci\theta}\times \\
&\quad\Big(\acoeff(\xt,\kw-\pw-\pw',\pw',\kw-\pw)\Wigner_0(\xt,\kw-\pw) \\
&\quad-\acoeff(\xt,\kw-\pw,\pw',\kw-\pw-\pw')\Wigner_0(\xt,\kw-\pw-\pw')\Big)\,.
\end{split}
\end{equation*}
Now because of \eref{def-aver}, we see that the average of $I_1$ will give rise to the Dirac factor $\dir(\pw+\pw')$. Therefore one introduces the notation:
\begin{displaymath}
\esp{\acoeff(\xt,\kw,\pw,\kw_1)\acoeff(\xt,\kw',\pw',\kw'_1)}=(2\pi)^8\dir(\pw+\pw')\aacoeff(\xt,\pw;\kw,\kw_1,\kw',\kw'_1)\,,
\end{displaymath}
and assumes that:
\begin{equation}\label{eq:mixing}
\esp{\smash{\TF{\varphi}_1(\pw)\otimes\TF{\psi}_1(\pw')}\Wigner_0(\xt,\kw)}=\esp{\smash{\TF{\varphi}_1(\pw)\otimes\TF{\psi}_1(\pw')}} \esp{\Wigner_0(\xt,\kw)}
\end{equation}
for $\smash{\varphi_1,\psi_1}\in\{\smash{\cper_1},\smash{\Vg_1}\}$, since the quantities $\smash{\varphi_1,\psi_1}$ and $\smash{\Wigner_0}$ vary on different scales. \alert{Indeed, $\smash{\cper_1}$ and $\smash{\Vg_1}$ are given in the definition \eqref{ela-ten-fluc} of the perturbations of the ambient flow, and $\Wigner_0$ is from the two-scale expansion \eqref{asym-expan}}. This crucial mixing assumption is also the one invoked in~\cite{RYZ96,BAL05}. Thus considering the change of variable $\smash{\pw+\pw'}\rightarrow\bzero$, one arrives at:
\begin{multline*}
\esp{I_1}=\demi\int_{\Rset^4} \frac{\aacoeff(\xt,\pw;\kw,\kw-\pw,\kw-\pw,\kw)\esp{\Wigner_0(\xt,\kw)}}{\vH(\xt,\kw)-\vH(\xt,\kw-\pw)+\ci\theta} \dd\pw \\
-\demi\int_{\Rset^4} \frac{\aacoeff(\xt,\pw;\kw,\kw-\pw,\kw,\kw-\pw)\esp{\Wigner_0(\xt,\kw-\pw)}}{\vH(\xt,\kw)-\vH(\xt,\kw-\pw)+\ci\theta} \dd\pw
\end{multline*}
because $\Wigner_0$ \emphes{does not depend on the fast scale}.

As for the last terms in \eref{second-correc-1}, namely:
\begin{multline*}
I_2=\Big[\vC(\xt,\kw-\Dx_\yt)\Vper(\yt)\cdot(\kg-\Dx_\yg)+\Vper(\yt)\cdot(\kg-\Dx_\yg)\vC(\xt,\kw-\Dx_\yt) \\
+\vC(\xt,\kw-\Dx_\yt)\cper_1(\yt)\vC(\xt,\kw-\Dx_\yt)\Big]\Wigner_1(\xt,\yt,\kw)\,,
\end{multline*}
observing that $\vC(\xt,\kw)$ is a (first-order) polynomial in $\kw$, one arrives by a straightforward direct computation using the definition (\ref{eq:PDO}):
\begin{equation*}
\begin{split}
I_2&=\int_{\Rset^8} \frac{\iexp^{\ci\yt\cdot(\pw+\pw')}\dd\pw\dd\pw'}{(2\pi)^8}\acoeff(\xt,\kw-\pw-\pw',\pw,\kw-\pw')\TF{\Wigner}_1(\xt,\pw',\kw) \\
&=\demi\int_{\Rset^8} \frac{\iexp^{\ci\yt\cdot(\pw+\pw')}\dd\pw\dd\pw'}{(2\pi)^8}\frac{\acoeff\left(\xt,\kw-\pw-\pw',\pw,\kw-\pw'\right)}{\vH(\xt,\kw)-\vH(\xt,\kw-\pw')-\ci\theta}\times \\
&\quad\Big(\acoeff(\xt,\kw-\pw',\pw',\kw)\Wigner_0(\xt,\kw)-\acoeff(\xt,\kw,\pw',\kw-\pw')\Wigner_0(\xt,\kw-\pw')\Big)\,.
\end{split}
\end{equation*}
Therefore:
\begin{multline*}
\esp{I_2}=-\demi\int_{\Rset^4} \frac{\aacoeff(\xt,-\pw;\kw,\kw-\pw,\kw-\pw,\kw)\esp{\Wigner_0(\xt,\kw)}}{\vH(\xt,\kw-\pw)-\vH(\xt,\kw)+\ci\theta} \dd\pw \\
+\demi\int_{\Rset^4} \frac{\aacoeff(\xt,-\pw;\kw,\kw-\pw,\kw,\kw-\pw)\esp{\Wigner_0(\xt,\kw-\pw)}}{\vH(\xt,\kw-\pw)-\vH(\xt,\kw)+\ci\theta} \dd\pw \,.
\end{multline*}
          
        
\subsubsection{Radiative transfer equation}
      
To conclude we insert in the average of \eref{second-correc-1} the foregoing expressions of $\smash{\esp{I_1}}$ and $\smash{\esp{I_2}}$. By a proper change of variable $\kw-\pw\rightarrow\pw$ in $\smash{I_1}$ and $\smash{I_2}$ it is deduced that:
\begin{equation}\label{eq:RTE-step00}
\begin{split}
\{\vL_0,\esp{\Wigner_0}\}=&-\frac{\apar(\xt)}{4}\int_{\Rset^4} \frac{\aacoeff(\xt,\kw-\pw;\kw,\pw,\pw,\kw)\esp{\Wigner_0(\xt,\kw)}}{\ci(\vH(\xt,\pw)-\vH(\xt,\kw))+\theta} \dd\pw \\
&+\frac{\apar(\xt)}{4}\int_{\Rset^4} \frac{\aacoeff(\xt,\kw-\pw;\kw,\pw,\kw,\pw)\esp{\Wigner_0(\xt,\pw)}}{\ci(\vH(\xt,\pw)-\vH(\xt,\kw))+\theta}  \dd\pw \\
&-\frac{\apar(\xt)}{4}\int_{\Rset^4} \frac{\aacoeff(\xt,\pw-\kw;\kw,\pw,\pw,\kw)\esp{\Wigner_0(\xt,\kw)}}{\ci(\vH(\xt,\kw)-\vH(\xt,\pw))+\theta} \dd\pw \\
&+\frac{\apar(\xt)}{4}\int_{\Rset^4} \frac{\aacoeff(\xt,\pw-\kw;\kw,\pw,\kw,\pw)\esp{\Wigner_0(\xt,\pw)}}{\ci(\vH(\xt,\kw)-\vH(\xt,\pw))+\theta}  \dd\pw \,.
\end{split}
\end{equation}
But $\vH(\xt,\kw)=0$ on the support of $\smash{\esp{\Wigner_0(\xt,\kw)}}$, and $\vH(\xt,\pw)=0$ on the support of $\smash{\esp{\Wigner_0(\xt,\pw)}}$. Also $\smash{\aacoeff(\xt,\kw-\pw;\cdot)}=\smash{\aacoeff(\xt,\pw-\kw;\cdot)}$ from the symmetry of the spectral density matrix, $\smash{\TF{\coro}(\kw-\pw)}=\smash{\TF{\coro}(\pw-\kw)^\itr}$. Therefore \eref{eq:RTE-step00} reduces to:
\begin{multline}\label{eq:RTE-step0}
\{\vL_0,\esp{\Wigner_0}\} = \\
\frac{\apar(\xt)}{2}\int_{\Rset^4} \left(\frac{\theta}{\vH(\xt,\kw)^2+\theta^2}\right)\aacoeff(\xt,\kw-\pw;\kw,\pw,\kw,\pw)\esp{\Wigner_0(\xt,\pw)}  \dd\pw \\
-\frac{\apar(\xt)}{2}\int_{\Rset^4} \left(\frac{\theta}{\vH(\xt,\pw)^2+\theta^2}\right)\aacoeff(\xt,\kw-\pw;\kw,\pw,\pw,\kw)\esp{\Wigner_0(\xt,\kw)}  \dd\pw\,.
\end{multline}

On the other hand, by letting \emphes{$\theta\rightarrow\smash{0^\pm}$} we have in the sense of distribution for $(\xt,\kw)\in\varXset$:
\begin{displaymath}
\frac{\theta}{\vH(\xt,\kw)^2+\theta^2}\rightarrow \pi\sig{\theta}\normu{\frac{\partial\vH}{\partial\wg}}^{-1}\left[\dir(\wg-\eigl_+(\xt,\kg))+\dir(\wg-\eigl_-(\xt,\kg))\right]\,,
\end{displaymath}
where $\smash{\sig{\theta}}$ stands for the sign of $\theta$, and $\smash{\normu{\frac{\partial\vH}{\partial\wg}}}=\norm{\vC(\xt,\kw)}=\Cref(\xt)\norm{\kg}$ whenever $\wg=\smash{\eigl_\pm(\xt,\kg)}$. A similar expression holds for $\smash{\frac{\theta}{\vH(\xt,\pw)^2+\theta^2}}$. In view of \eref{eq:def-alpha}, one then introduces the notation:
\begin{equation}\label{eq:dscat}
\dscatij_d(\xt;\kw|\pw) = \frac{2\pi}{\Cref^2(\xt)\norm{\kg}\norm{\pg}}{\boldsymbol T}(\xt,\kw,\pw)^\itr\TF{\coro}(\kw-\pw){\boldsymbol T}(\xt,\kw,\pw)\,,
\end{equation}
where:
\begin{displaymath}
{\boldsymbol T}(\xt,\kw,\pw)=\demi\begin{pmatrix} \vC(\kw)\vC(\pw) \\ \vC(\kw+\pw)\pg \\  \end{pmatrix}\,,
\end{displaymath}
and observe that $\kw=(\kg,\wg)$ and $\pw=(\pg,\upsilon)$ do not \emph{a priori} play symmetric roles in this expression: $\smash{\dscatij_d(\xt;\kw|\pw)}\neq\smash{\dscatij_d(\xt;\pw|\kw)}$. One also defines:
\begin{equation}
\dscatij_t(\xt;\pw|\kw) = \frac{2\pi}{\Cref^2(\xt)\norm{\kg}\norm{\pg}}{\boldsymbol T}(\xt,\kw,\pw)^\itr\TF{\coro}(\kw-\pw){\boldsymbol T}(\xt,\pw,\kw)\,,
\end{equation}
and:
\begin{equation}\label{eq:tscat}
\tscati_t(\xt,\kw) = \int_{\Rset^3}\left(\dscatij_t(\xt;\pg,\eigl_-(\xt,\pg)|\kw)+\dscatij_t(\xt;\pg,\eigl_+(\xt,\pg)|\kw)\right)\dd\pg\,.
\end{equation}
Chosing $\sig{\theta}=\sig{\vC}$ (the sign of $\vC\neq0$ on $\varXset$) to preserve causality \cite{BAL05,RYZ96} and recalling that $\vL_0(\xt,\kw)=\apar(\xt)\vH(\xt,\kw)$, \eref{eq:RTE-step0} finally reads as the following radiative transfer equation:
\begin{multline}\label{eq:RTE}
\{\vH,\apar\esp{\Wigner_0}\}+\sig{\vC}\tscati_t(\xt,\kw)\apar(\xt)\Cref(\xt)\norm{\kg}\esp{\Wigner_0(\xt,\kw)}= \\
\sig{\vC}\int_{\Rset^4}\left(\dscatij_d(\xt;\kg,\eigl_-(\xt,\kg)|\pw)+\dscatij_d(\xt;\kg,\eigl_+(\xt,\kg)|\pw)\right)\apar(\xt)\Cref(\xt)\norm{\pg}\esp{\Wigner_0(\xt,\pw)}\dd\pw
\end{multline}
This is \eref{eq:RTE-acout} of the introduction, where one recognizes the average wave action $\action=\apar\norm{\vC}\esp{\Wigner_0}$ with $\smash{\vC(\xt,\kg,\eigl_\pm(\xt,\kg))}=\mp\smash{\Cref(\xt)\norm{\kg}}$ on the support of $\smash{\esp{\Wigner_0}}$ (up to the proper redefinition $\smash{\lambda_\pm}=-\smash{\eigl_\pm}$). It describes the evolution of the wave action in a randomly perturbed, inhomogeneous ambient flow. It is very much similar to the kinetic equations for the spectrum of the Fourier coefficients of the density disturbances derived by Howe by an alternative approach, starting from Lighthill's acoustic analogy; see \cite[Eqs. (37) and (39)]{HOW73}. The kernel $\smash{\dscatij_d}(\xt;\kw|\pw)$ is the rate of conversion of energy with wave vector and circular frequency $\pw$ into energy with wave vector and frequency $\kw$, at position and time $\xt$, and is called a differential scattering cross-section. Parallely, $\smash{\tscati_t(\xt,\kw)}$ is the total scattering cross-section accounting for all conversions into energy with wave vector and frequency different from $\kw$ at position and time $\xt$.

Now proceeding as for \eref{eq:Liouville-equation} in view of \eref{dis-rel-3} and with the notation (\ref{eq:action}), the foregoing radiative transfer equation also reads:
\begin{multline}\label{eq:RTE-action}
\partial_t\action_+(\xt,\kg)+\vray^+\cdot\bnabla_\xg\action_+-\bnabla_\xg(\vray^+\cdot\kg)\cdot\bnabla_\kg\action_++\tscati_+(\xt,\kg)\action_+(\xt,\kg) \\
=\int_{\Rset^3}\left(\dscatij_{++}(\xt;\kg|\pg)\action_+(\xt,\pg)+\dscatij_{+-}(\xt;\kg|\pg)\action_-(\xt,\pg)\right)\dd\pg\,, \\
\partial_t\action_-(\xt,\kg)+\vray^-\cdot\bnabla_\xg\action_--\bnabla_\xg(\vray^-\cdot\kg)\cdot\bnabla_\kg\action_-+\tscati_-(\xt,\kg)\action_-(\xt,\kg) \\
=\int_{\Rset^3}\left(\dscatij_{-+}(\xt;\kg|\pg)\action_+(\xt,\pg)+\dscatij_{--}(\xt;\kg|\pg)\action_-(\xt,\pg)\right)\dd\pg\,, \
\end{multline}
where the differential and total scattering cross-sections are now:
\begin{equation*}
\begin{split}
\dscatij_{\pm\pm}(\xt;\kg|\pg) &=\dscatij_d(\xt;\kg,\eigl_\pm(\xt,\kg)|\pg,\eigl_\pm(\xt,\pg))\,,
\end{split}
\end{equation*}
and:
\begin{equation*}
\tscati_\pm(\xt,\kg)= \int_{\Rset^3}\left(\dscatij_t(\xt;\pg,\eigl_-(\xt,\pg)|\kg,\eigl_\pm(\xt,\kg))+\dscatij_t(\xt;\pg,\eigl_+(\xt,\pg)|\kg,\eigl_\pm(\xt,\kg)\right)\dd\pg\,.
\end{equation*}
If one neglects for example the correlation of the perturbations of the speed of sound and the particle velocity ($\smash{\TF{\coro}_{cv}}(\kw-\pw)=\bzero$), these expressions are:
\begin{equation}
\begin{split}
\dscatij_{++}(\xt;\kg|\pg)= &\frac{\pi}{2}\Cref^2(\xt)\norm{\kg}\norm{\pg}\TF{\coroij}_c(\kg-\pg,\eigl_+(\kg)-\eigl_+(\pg)) \\
&+\frac{\pi}{2}\frac{(\norm{\kg}+\norm{\pg})^2}{\norm{\kg}\norm{\pg}}\pg\cdot\TF{\coro}_v(\kg-\pg,\eigl_+(\kg)-\eigl_+(\pg))\pg\,, \\
\dscatij_{+-}(\xt;\kg|\pg)= &\frac{\pi}{2}\Cref^2(\xt)\norm{\kg}\norm{\pg}\TF{\coroij}_c(\kg-\pg,\eigl_+(\kg)-\eigl_-(\pg)) \\
&+\frac{\pi}{2}\frac{(\norm{\kg}-\norm{\pg})^2}{\norm{\kg}\norm{\pg}}\pg\cdot\TF{\coro}_v(\kg-\pg,\eigl_+(\kg)-\eigl_-(\pg))\pg\,,
\end{split}
\end{equation}
and:
\begin{equation}
\begin{split}
\tscati_+(\xt,\kg)=\frac{\pi}{2}\int_{\Rset^3}&\Big[\Cref^2(\xt)\norm{\kg}\norm{\pg}\TF{\coroij}_c(\kg-\pg,\eigl_+(\kg)-\eigl_+(\pg)) \\
&+\Cref^2(\xt)\norm{\kg}\norm{\pg}\TF{\coroij}_c(\kg-\pg,\eigl_+(\kg)-\eigl_+(\pg)) \\
&+(\norm{\kg}+\norm{\pg})^2\hpg\cdot\TF{\coro}_v(\kg-\pg,\eigl_+(\kg)-\eigl_-(\pg))\hkg \\
&+(\norm{\kg}-\norm{\pg})^2\hpg\cdot\TF{\coro}_v(\kg-\pg,\eigl_+(\kg)-\eigl_-(\pg))\hkg\Big] \dd\pg\,.
\end{split}
\end{equation}
If the perturbations of the particle velocity are in addition divergence-free, it can be readily checked that \cite{FAN01}:
\begin{displaymath}
\pg\cdot\TF{\coro}_v(\kg-\pg,\wg)\pg=\pg\cdot\TF{\coro}_v(\kg-\pg,\wg)\kg=\kg\cdot\TF{\coro}_v(\kg-\pg,\wg)\kg=\kg=\kg\cdot\TF{\coro}_v(\kg-\pg,\wg)\pg\,.
\end{displaymath}
Therefore $\dscatij_d(\xt;\kw|\pw)=\dscatij_d(\xt;\pw|\kw)=\dscatij_t(\xt;\kw|\pw)=\dscatij_t(\xt;\pw|\kw)$ and:
\begin{displaymath}
\begin{split}
\tscati_+(\xt,\kg) &=\int\left(\dscatij_{++}(\xt;\kg|\pg)+\dscatij_{+-}(\xt;\kg|\pg)\right)\dd\pg\,, \\
\tscati_-(\xt,\kg) &=\int\left(\dscatij_{-+}(\xt;\kg|\pg)+\dscatij_{--}(\xt;\kg|\pg)\right)\dd\pg\,.
\end{split}
\end{displaymath}
Thus the equations of radiative transfer (\ref{eq:RTE-action}) are conservative for the overall wave action $\action=\action_+\otimes\dir(\wg-\eigl_+)+\action_-\otimes\dir(\wg-\eigl_-)$: \emphes{$\iint(\action_++\action_-)\dd\xg\dd\kg$ is constant, ignoring however possible boundary effects in $\domain\times\smash{\Rset^3}$ (see \cite{AKI12} for the consideration of the hyperbolic set)}. This result is to be paralleled with the conclusions of \cite{FAN01}, where the flow-acoustic scattering is shown to become non conservative because of the flow-straining term (non vanishing gradient of the ambient quantities).

%

\subsection{Particular case: quiescent ambient medium}

We consider the case of a quiescent medium with $\vref(\xg,t)=\bzero$ and a sound speed $\cref(\xg)$ and density $\rref(\xg)$ which are independent of time. Then \eref{eq:conv-wave} reads:
\begin{equation}\label{eq:acou-wave}
\frac{1}{\rref}\bnabla_\xg\cdot(\rref\bnabla_\xg\vpot)-\frac{1}{\cref^2}\frac{\partial^2\vpot}{\partial t^2}=0\,,
\end{equation}
to be compared with the classical acoustic wave equation for the pressure field in an inhomogeneous medium \cite{BER46}:
\begin{displaymath}
\rref\bnabla_\xg\cdot\left(\frac{1}{\rref}\bnabla_\xg\pres'\right)-\frac{1}{\cref^2}\frac{\partial^2\pres'}{\partial t^2}=0\,.
\end{displaymath}
Consequently $\vH(\xg;\kg,\wg)=\smash{\demi(\wg^2-\Cref^2(\xg)\norm{\kg}^2)}$ is independent of time and accounting for random perturbations of the speed of sound solely one has:
\begin{displaymath}
\TF{\coroij}_c(\kg,\wg)=\TF{{\mathcal R}}_c(\kg)\otimes\dir(\wg)
\end{displaymath}
for the correlation time is "infinite" in this situation. Thus $\smash{\eigl_\pm(\xg,\kg)}=\mp\smash{\Cref(\xg)\norm{\kg}}$ and one arrives at the radiative transfer equation for \emphes{$\action_\pm(\xg,t;\kg)$ ($\action_+$ and $\action_-$ being uncoupled in this case)}:
\begin{multline}\label{eq:RTE-acoubis}
\partial_t\action_\pm \pm\Cref(\xg)\hkg\cdot\bnabla_\xg\action_\pm \mp\norm{\kg}\bnabla_\xg\Cref\cdot\bnabla\action_\pm= \\
\int_{\Rset^3}\dscatij(\xg;\kg|\pg)\left(\action_\pm(\xg,t;\pg)-\action_\pm(\xg,t;\kg)\right)\dir(\Cref(\xg)\norm{\pg}-\Cref(\xg)\norm{\kg})\dd\pg\,,
\end{multline}
where the scattering cross-section is:
\begin{equation}\label{eq:dscat-acou}
\dscatij(\xg;\kg|\pg) = \frac{\pi}{2}\Cref^2(\xg)\norm{\kg}^2\TF{{\mathcal R}}_c(\kg-\pg)\,.
\end{equation}
This result agrees with the models developed in \cite{BAL05,RYZ96}. \eref{eq:RTE-acoubis} is the counterpart of \eref{eq:RTE-acou} of the introduction for the energy density itself. In this situation, even if all waves have the same frequency, they may nevertheless interact if they have different wave numbers. The right-hand side of \eref{eq:RTE-acoubis} describes this wave number conversion process. 

\section{Conclusions}\label{sec:CL}

In this paper, a (kinetic) radiative transfer equation describing the propagation of high-frequency acoustic waves in arbitrarily random ambient flows has been derived. The model accounts for possible perturbations of the ambient particle velocity and ambient speed of sound when both quantities vary spatially and temporally. These situations correspond to random variations of the temperature (and humidity in the atmosphere) and random variations of the possibly turbulent particle velocity induced by the presence of sound waves, at the small scale (wavelength) of the acoustic waves. In this respect, it generalizes the previous results established for high-frequency acoustic waves in a quiescent medium \cite{BAL05,RYZ96} and in a frozen ambient flow \cite{FAN01}. The proposed model also extends earlier works for time-varying ambient flows \cite{HOW73} where radiative transfer equations were obtained by a formal approach for specific forms of the perturbations of the ambient quantities. The links with classical geometric acoustics are also established. We have used a Wigner functional approach to derive the radiative transfer equation which basically describes the evolution of the angularly and frequency resolved action of the acoustic waves in phase space. It also describes the phase shift, spectral broadening, and multiple scattering effects. In future works we should consider the diffusion limit of the radiative transfer model, whereby the wave action is evolved in physical space solely thus reducing the dimension of the kinetic equation. \alert{Another issue for practical applications is the derivation of \emph{ad hoc} boundary conditions for the Wigner measures, which raises challenging mathematical questions. They are addressed in \cite{AKI12} for the case of transverse reflections in quiescent media, however further investigations are needed for the cases of tangent and total reflections in quiescent or moving media}.



\newpage

\appendix

\section{Formal self-adjointness of the convected wave equation (\ref{eq:conv-wave})}\label{apx:proof-transport}

\newcommand{\scaldd}[1]{\left(\left(#1\right)\right)}

\noindent We define $\scaldd{\fu,\fv}$ for the $\smash{L^2(\Rset^3_\xg\times\Rset_t,\Cset)}$ scalar product. Then $\forall\fu,\fv\in\smash{\varCset^\infty(\Rset^3_\xg\times\Rset_t)}$ with either $\fu$ or $\fv$ being compactly supported:
\begin{displaymath}
\begin{split}
\scaldd{\bnabla_\xg\cdot(\rref\bnabla_\xg\fu),\fv} &=-\scaldd{\rref\bnabla_\xg\fu,\bnabla_\xg\fv}=-\scaldd{\bnabla_\xg\fu,\rref\bnabla_\xg\fv} \\
&=\scaldd{\fu,\bnabla_\xg\cdot(\rref\bnabla_\xg\fv)}\,.
\end{split}
\end{displaymath}
\noindent Likewise:
\begin{displaymath}
\begin{split}
\scaldd{\dconv{\fu},\fv} &=\scaldd{\partial_t\fu+\vref\cdot\bnabla_\xg\fu,\fv} \\
&=-\scaldd{\fu,\partial_t\fv+\bnabla_\xg\cdot(\fv\vref)} \\
&=-\scaldd{\fu,\dconv{\fv}+\fv\bnabla_\xg\cdot\vref}\,,
\end{split}
\end{displaymath}
so that the formal adjoint of the convective derivative (\ref{eq:dconv0}) is $\smash{\adj{\dconv{\fu}}}=-\smash{\dconv{\fu}-\fu\bnabla_\xg\cdot\vref}$. Therefore:
\begin{displaymath}
\begin{split}
\scaldd{\rref\dconv{}\left(\frac{1}{\cref^2}\dconv{\fu}\right),\fv} &=\scaldd{\dconv{}\left(\frac{1}{\cref^2}\dconv{\fu}\right),\rref\fv} \\
&=-\scaldd{\frac{1}{\cref^2}\dconv{\fu},\dconv{(\rref\fv)}+\rref\fv\bnabla_\xg\cdot\vref} \\
&=-\scaldd{\frac{1}{\cref^2}\dconv{\fu},\fv\cancel{\left(\dconv{\rref}+\rref\bnabla_\xg\cdot\vref\right)}+\rref\dconv{\fv}} \\
&=-\scaldd{\dconv{\fu},\frac{\rref}{\cref^2}\dconv{\fv}} \\
&=-\cjg{\scaldd{\dconv{\fv},\frac{\rref}{\cref^2}\dconv{\fu}}} \\
&=\cjg{\scaldd{\rref\dconv{}\left(\frac{1}{\cref^2}\dconv{\fv}\right),\fu}} \\
&=\scaldd{\fu,\rref\dconv{}\left(\frac{1}{\cref^2}\dconv{\fv}\right)}
\end{split}
\end{displaymath}
where the conjugate equalities stem from the fact that $\fu$ and $\fv$ play symmetric roles; also we have used the mass-conservation equation (\ref{eq:ambient-cons}). We can thus conclude that the convected wave equation (\ref{eq:conv-wave}) is formally self-adjoint. Multiplying it by $\smash{(\frac{\escale}{\ci})^2}$ we have for all compactly supported, smooth test function $\fv$:
\begin{displaymath}
\scaldd{\vL_\escale(\xg,t,\escale\Dxx,\escale\Dt)\fu,\fv}=\scaldd{\fu,\vL_\escale(\xg,t,\escale\Dxx,\escale\Dt)\fv}\,.
\end{displaymath}

\section{Proof of the rules (\ref{reg1-cal}) and (\ref{reg2-cal})}\label{apx:proof1}

\newcommand{\eDt}{\escale\iD_t}
\newcommand{\eDx}{\escale\Dxx}

Invoking the trace formula (\ref{eq:trace}) one has for any smooth function $\smash{\obsj_1\in\varCset_0^\infty(\Rset^n_\xg\times\Rset_\kg^n)}$:
\begin{equation}\label{eq:dir1}
\scal{\Wignere[\obsj_2(\xg,\eDx)\fu,\fv],\obsj_1}=\scald{\obsj_1(\xg,\eDx)\obsj_2(\xg,\eDx)\fu,\fv}\,,
\end{equation}
where $\scal{\cdot,\cdot}$ denotes the duality bracket between $\varSset$ and $\varSset'$ on $\Rset_\xg^n\times\Rset_\kg^n$. But according to \cite[Th. 2.7.4]{MAR02} the operator $\obsj_1(\xg,\eDx)\obsj_2(\xg,\eDx)$ can be identified with the operator $\obsj(\xg,\eDx)$ provided that:
\begin{equation}\label{eq:productPDO}
\obsj(\xg,\kg)=\obsj_1(\xg,\kg)\obsj_2(\xg,\kg)+\frac{\escale}{\ci}\bnabla_\kg\obsj_1\cdot\bnabla_\xg\obsj_2+\Go(\escale^2)\,,
\end{equation}
where all higher-order terms are explicitly given in \cite[Ch. 6]{MAR02}. Therefore:
\begin{equation}\label{eq:dir2}
\begin{split}
\scald{\obsj_1(\xg,\eDx)\obsj_2(\xg,\eDx)\fu,\fv} &=\scal{\Wignere[\fu,\fv],\obsj_1\obsj_2} \\
&\quad+\frac{\escale}{\ci}\scal{\Wignere[\fu,\fv],\bnabla_\kg\obsj_1\cdot\bnabla_\xg\obsj_2}+\Go(\escale^2) \\
&=\scal{\obsj_2\Wignere[\fu,\fv],\obsj_1}-\frac{\escale}{\ci}\scal{\bnabla_\xg\obsj_2\cdot\bnabla_\kg\Wignere[\fu,\fv],\obsj_1} \\
&\quad-\frac{\escale}{\ci}\scal{(\bnabla_\xg\cdot\bnabla_\kg\obsj_2)\Wignere[\fu,\fv],\obsj_1}+\Go(\escale^2)\,.
\end{split}
\end{equation}
Identifying (\ref{eq:dir1}) and (\ref{eq:dir2}) one formally obtains the rule (\ref{reg1-cal}). As for the adjoint state, one has invoking again the trace formula (\ref{eq:trace}):
\begin{equation}\label{eq:cjg1}
\begin{split}
\scal{\Wignere[\fu,\obsj_2(\xg,\eDx)\fv],\obsj_1} &=\scald{\obsj_1(\xg,\eDx)\fu,\obsj_2(\xg,\eDx)\fv} \\
&=\scald{\adj{\obsj_2(\xg,\eDx)}\obsj_1(\xg,\eDx)\fu,\fv}\,.
\end{split}
\end{equation}
But according to \cite[Rem. 2.5.7]{MAR02} the operator $\adj{\obsj(\xg,\eDx)}$ can be identified with the operator $\tilde{\obsj}(\xg,\eDx)$ provided that:
\begin{equation}\label{eq:adjPDO}
\tilde{\obsj}(\xg,\kg)=\adj{\obsj}(\xg,\kg)+\frac{\escale}{\ci}\bnabla_\kg\cdot\bnabla_\xg\adj{\obsj}+\Go(\escale^2)\,,
\end{equation}
where all higher-order terms are explicitly given in \cite[Ch. 6]{MAR02}. Therefore combining \eref{eq:productPDO} and \eref{eq:adjPDO} one has:
\begin{equation}\label{eq:cjg2}
\begin{split}
\scald{\adj{\obsj_2(\xg,\eDx)}\obsj_1(\xg,\eDx)\fu,\fv} &=\scal{\Wignere[\fu,\fv],\adj{\obsj}_2\obsj_1}+\frac{\escale}{\ci}\scal{\Wignere[\fu,\fv],(\bnabla_\kg\cdot\bnabla_\xg\adj{\obsj}_2)\obsj_1} \\
&\quad+\frac{\escale}{\ci}\scal{\Wignere[\fu,\fv],\bnabla_\kg\adj{\obsj}_2\cdot\bnabla_\xg\obsj_1} + \Go(\escale^2) \\
&=\scal{\adj{\obsj}_2\Wignere[\fu,\fv],\obsj_1} \\
&\quad-\frac{\escale}{\ci}\scal{\bnabla_\kg\adj{\obsj}_2\cdot\bnabla_\xg\Wignere[\fu,\fv],\obsj_1} + \Go(\escale^2)\,.
\end{split}
\end{equation}
Identifying (\ref{eq:cjg1}) and (\ref{eq:cjg2}) one formally obtains the rule (\ref{reg2-cal}). Additionally if $\smash{\obsj_2(\xg,\eDx)}$ is formally self-adjoint, it turns out that:
\begin{displaymath}
\begin{split}
\scal{\Wignere[\fu,\obsj_2(\xg,\eDx)\fv],\obsj_1} &=\scald{\obsj_2(\xg,\eDx)\obsj_1(\xg,\eDx)\fu,\fv} \\
&=\scal{\Wignere[\fu,\fv],\obsj_2\obsj_1}+\frac{\escale}{\ci}\scal{\Wignere[\fu,\fv],\bnabla_\kg\obsj_2\cdot\bnabla_\xg\obsj_1} + \Go(\escale^2) \\
&=\scal{\obsj_2\Wignere[\fu,\fv],\obsj_1}-\frac{\escale}{\ci}\scal{\bnabla_\kg\obsj_2\cdot\bnabla_\xg\Wignere[\fu,\fv],\obsj_1} \\
&\quad-\frac{\escale}{\ci}\scal{(\bnabla_\kg\cdot\bnabla_\xg\obsj_2)\Wignere[\fu,\fv],\obsj_1}+\Go(\escale^2)\,.
\end{split}
\end{displaymath}
Therefore, one formally has in this situation:
\begin{equation}\label{eq:selfadj}
\Wignere[\obsj_2(\xg,\eDx)\fu,\fv]-\Wignere[\fu,\obsj_2(\xg,\eDx)\fv]=\frac{\escale}{\ci}\{\obsj_2,\Wignere[\fu,\fv]\}+\Go(\escale^2)\,,
\end{equation}
which is \eref{reg3-cal}.

Besides, we explicitely have:
\begin{multline*}
\scal{\Wignere[\fu,\obsj_2(\xg,\eDx)\fv],\obsj_1} \\
=\int\dd\xg\iint\frac{\dd\kg\dd\yg}{(2\pi)^n}\iexp^{\ci\kg\cdot(\xg-\yg)}\obsj_1(\xg,\escale\kg)\fu(\yg)\iint\frac{\dd\pg\dd\yg'}{(2\pi)^n}\iexp^{-\ci\pg\cdot(\xg-\yg')}\adj{\obsj}_2(\xg,\escale\pg)\adj{\fv}(\yg') \\
=\iint\frac{\escale^n\dd\kg\dd\pg}{(2\pi)^n}\iiint\frac{\dd\xg\dd\yg\dd\yg'}{(2\pi)^n}\iexp^{\ci(\kg-\pg)\cdot(\xg-\yg')+\ci\escale\kg\cdot\yg}\obsj_1(\xg,\escale\kg)\adj{\obsj}_2(\xg,\escale\pg)\fu(\yg'-\escale\yg)\adj{\fv}(\yg')
\end{multline*}
by the change of variable $\yg\rightarrow\yg'-\escale\yg$. Now by the change of variable $\escale\kg\rightarrow\kg$ we arrive at:
\begin{multline*}
\scal{\Wignere[\fu,\obsj_2(\xg,\eDx)\fv],\obsj_1} \\
=\iint\frac{\dd\kg\dd\pg}{(2\pi)^n}\iint\dd\xg\dd\yg'\iexp^{\ci(\frac{\kg}{\escale}-\pg)\cdot(\xg-\yg')}\obsj_1(\xg,\kg)\adj{\obsj}_2(\xg,\escale\pg)\int\frac{\dd\yg}{(2\pi)^n}\iexp^{\ci\kg\cdot\yg}\fu(\yg'-\escale\yg)\adj{\fv}(\yg') \\
=\iint\frac{\dd\kg\dd\pg}{(2\pi)^n}\iint\dd\xg\dd\yg'\iexp^{\ci(\frac{\kg}{\escale}-\pg)\cdot(\xg-\yg')}\obsj_1(\xg,\kg)\adj{\obsj}_2(\xg,\escale\pg)\Wignere[\fu,\fv](\yg',\kg)\,.
\end{multline*}
The additional changes of variable $\vp\rightarrow\smash{\frac{\vp}{\escale}}$ and $\yg'\rightarrow\xg-\escale\yg'$ yield:
\begin{multline*}
\scal{\Wignere[\fu,\obsj_2(\xg,\eDx)\fv],\obsj_1} \\
=\iint\frac{\dd\kg\dd\pg}{(2\pi)^n}\iint\dd\xg\dd\yg'\iexp^{\ci(\kg-\pg)\cdot\yg'}\obsj_1(\xg,\kg)\adj{\obsj}_2(\xg,\pg)\Wignere[\fu,\fv](\xg-\escale\yg',\kg)\,,
\end{multline*}
so that:
\begin{displaymath}
\Wignere[\fu,\obsj_2(\xg,\eDx)\fv] (\xg,\kg)=\iint\frac{\dd\yg'\dd\pg}{(2\pi)^n}\iexp^{\ci(\kg-\pg)\cdot\yg'}\adj{\obsj}_2(\xg,\pg)\Wignere[\fu,\fv](\xg-\escale\yg',\kg)\,.
\end{displaymath}
However:
\begin{displaymath}
\int_{\Rset^n}\iexp^{\ci\yg\cdot\kg}f(\xg-\escale\yg)\dd\yg = \escale^{-n}\iexp^{\ci\xg\cdot\frac{\kg}{\escale}}\TF{f}\left(\frac{\kg}{\escale}\right)\,,
\end{displaymath}
so that we finally arrive at:
\begin{displaymath}
\begin{split}
\Wignere[\fu,\obsj_2(\xg,\eDx)\fv] (\xg,\kg) &=\int_{\Rset^n}\frac{\dd\pg}{(2\pi\escale)^n}\iexp^{\ci\xg\cdot\frac{(\kg-\pg)}{\escale}}\adj{\obsj}_2(\xg,\pg)\TF{\Wigner}_\escale[\fu,\fv]\left(\frac{\kg-\vp}{\escale},\kg\right) \\
&=\int_{\Rset^n}\frac{\dd\pg}{(2\pi)^n}\iexp^{\ci\xg\cdot\vp}\adj{\obsj}_2(\xg,\kg-\escale\pg)\TF{\Wigner}_\escale[\fu,\fv](\vp,\kg) \\
&=\left(\adj{\obsj}_2(\xg,\kg-\escale \Dxx)\Wignere[\fu,\fv]\right)(\xg,\kg)\,,
\end{split}
\end{displaymath}
which is the claimed formula (\ref{reg2-cal}).

\section{Proof of the rules (\ref{reg5-cal}) and (\ref{reg5adj-cal})}\label{apx:proof2}

\newcommand{\xgep}{\frac{\xg}{\escale}}

We start by showing how the rules (\ref{reg4-cal}) arise. Invoking once again the trace formula (\ref{eq:trace}) one has for any smooth function $\smash{\obsj\in\varCset_0^\infty(\Rset^n_\xg\times\Rset_\kg^n)}$:
\begin{equation}\label{eq:reg5-step0}
\scal{\Wignere\left[f\left(\xgep\right)\fu,\fv\right],\obsj}=\scald{\obsj(\xg,\eDx)f\left(\xgep\right)\fu,\fv}\,.
\end{equation}
But $\TF{f(\frac{\cdot}{\escale})}(\kg)=\escale^n\TF{f}(\escale\kg)$ such that:
\begin{displaymath}
\TF{f\left(\frac{\cdot}{\escale}\right)\fu(\cdot)}(\kg)=\int_{\Rset^n}\frac{\escale^n\dd\pg}{(2\pi)^n}\TF{f}(\escale\pg)\TF{\fu}(\kg-\pg)\,,
\end{displaymath}
and therefore:
\begin{displaymath}
\begin{split}
\obsj(\xg,\eDx)f\left(\xgep\right)\fu(\xg) &=\int_{\Rset^n}\frac{\dd\kg}{(2\pi)^n}\iexp^{\ci\xg\cdot\kg}\obsj(\xg,\escale\kg)\TF{f\left(\frac{\cdot}{\escale}\right)\fu(\cdot)}(\kg) \\
&=\int_{\Rset^n}\frac{\dd\kg}{(2\pi)^n}\iexp^{\ci\xg\cdot\kg}\obsj(\xg,\escale\kg)\int_{\Rset^n}\frac{\escale^n\dd\pg}{(2\pi)^n}\TF{f}(\escale\pg)\TF{\fu}(\kg-\pg) \\
&=\int_{\Rset^n}\frac{\dd\kg}{(2\pi\escale)^n}\iexp^{\ci\xgep\cdot\kg}\obsj(\xg,\kg)\int_{\Rset^n}\frac{\dd\pg}{(2\pi)^n}\TF{f}(\pg)\TF{\fu}\left(\frac{\kg-\pg}{\escale}\right)\,.
\end{split}
\end{displaymath}
However:
\begin{displaymath}
\TF{\fu}\left(\frac{\kg-\pg}{\escale}\right)=\int_{\Rset^n}\iexp^{-\ci\frac{\yg}{\escale}\cdot(\kg-\pg)}\fu(\yg)\dd\yg=\int_{\Rset^n}\iexp^{-\ci(\xgep-\yg)\cdot(\kg-\pg)}\fu(\xg-\escale\yg)\escale^n\dd\yg\,,
\end{displaymath}
so that one finally has:
\begin{multline*}
\scald{\obsj(\xg,\eDx)f\left(\xgep\right)\fu,\fv} \\
=\int_{\Rset^n}\frac{\dd\kg}{(2\pi\escale)^n}\iexp^{\ci\xgep\cdot\kg}\obsj(\xg,\kg)\int_{\Rset^n}\frac{\dd\pg}{(2\pi)^n}\TF{f}(\pg)\iint_{\Rset_\xg^n\times\Rset_\yg^n}\iexp^{-\ci(\xgep-\yg)\cdot(\kg-\pg)}\fu(\xg-\escale\yg)\cjg{\fv(\xg)}\escale^n\dd\yg\dd\xg \\
=\iint_{\Rset^n_\xg\times\Rset^n_\kg}\dd\xg\dd\kg\,\obsj(\xg,\kg)\int_{\Rset^n}\frac{\dd\pg}{(2\pi)^n}\iexp^{\ci\xgep\cdot\pg}\TF{f}(\pg)\Wignere[\fu,\fv](\xg,\kg-\pg)\,,
\end{multline*}
which when identified with \eref{eq:reg5-step0} gives the claimed result. Regarding \eref{reg5-cal}, it now suffices to observe that:
\begin{displaymath}
\begin{split}
\Wignere\left[f\left(\xgep\right)\obsj(\xg,\eDx)\fu,\fv\right] &=\int_{\Rset^n}\frac{\dd\pg}{(2\pi)^n}\iexp^{\ci\xgep\cdot\pg}\TF{f}(\pg)\Wignere[\obsj(\xg,\eDx)\fu,\fv](\xg,\kg-\pg) \\
&=\int_{\Rset^n}\frac{\dd\pg}{(2\pi)^n}\iexp^{\ci\xgep\cdot\pg}\TF{f}(\pg)\obsj(\xg,\kg-\pg)\Wignere[\fu,\fv](\xg,\kg-\pg) + \Go(\escale)\,,
\end{split}
\end{displaymath}
applying the rule (\ref{reg1-cal}).
 
\end{document}